\documentclass[apj,twocolumn]{openjournal}

\usepackage{graphicx}
\usepackage{xcolor}
\usepackage{MnSymbol}
\usepackage{multirow}
\usepackage{comment}
\usepackage{adjustbox}
\usepackage{rotating}
\usepackage[breaklinks,colorlinks,citecolor=blue,urlcolor=blue,linkcolor=blue]{hyperref}

\def\gtorder{\mathrel{\raise.3ex\hbox{$>$}\mkern-14mu
             \lower0.6ex\hbox{$\sim$}}}
\def\ltorder{\mathrel{\raise.3ex\hbox{$<$}\mkern-14mu
             \lower0.6ex\hbox{$\sim$}}}

\begin{document}

\title{ASASSN-24fw: An 8-month long, 4.1 mag, optically achromatic and polarized dimming event}

\author{\vspace{-1.3cm}R.~For\'es-Toribio\,$^{1,2}$}
\email[Email: ]{forestoribio.1@osu.edu}
\author{B.~JoHantgen\,$^{1}$}
\author{C.~S.~Kochanek\,$^{1,2}$}
\author{S.~G.~Jorstad\,$^{3}$}
\author{J.~J.~Hermes\,$^{4}$}
\author{J.~D.~Armstrong\,$^{5}$}
\author{C.~Ashall\,$^{6}$}
\author{C.~R.~Burns\,$^{7}$}
\author{E.~Gaidos\,$^{8,9}$}
\author{W.~B.~Hoogendam\,$^{6,*}$}
\altaffiliation{$^{*}$NSF Graduate Research Fellow}
\author{E.~Y.~Hsiao\,$^{10}$}
\author{K.~Medler\,$^{6}$}
\author{N.~Morrell\,$^{11}$}
\author{C.~Pfeffer\,$^{6}$}
\author{B.~J.~Shappee\,$^{6}$}
\author{K.~Stanek\,$^{1,2}$}
\author{M.~A.~Tucker\,$^{1,2}$}
\author{H.~Xiao\,$^{10}$}
\author{K. Auchettl\,$^{12,13}$}
\author{L.~Lu\,$^{1,2}$}
\author{D.~M.~Rowan\,$^{1,2}$}
\author{T.~Vaccaro\,$^{14}$}
\author{J.~P.~Williams\,$^{6}$}

\affiliation{$^{1}$Department of Astronomy, The Ohio State University, 140 West 18th Avenue, Columbus, OH 43210, USA}
\affiliation{$^{2}$Center for Cosmology and Astroparticle Physics, The Ohio State University, 191 W. Woodruff Avenue, Columbus, OH 43210, USA}
\affiliation{$^{3}$Institute for Astrophysical Research, Boston University, 725 Commonwealth Avenue, Boston, MA 02215, USA}
\affiliation{$^{4}$Department of Astronomy \& Institute for Astrophysical Research, Boston University, 725 Commonwealth Ave., Boston, MA 02215, USA}
\affiliation{$^{5}$University of Hawai'i Institute for Astronomy 34 Ohia Ku Street, Pukalani, HI 96768, USA }
\affiliation{$^{6}$Institute for Astronomy, University of Hawai'i at Manoa, 2680 Woodlawn Dr., Hawai'i, HI 96822, USA }
\affiliation{$^{7}$The Observatories of the Carnegie Institution for Science, 813 Santa Barbara St., Pasadena, CA 91101, USA}
\affiliation{$^{8}$Department of Earth Sciences, University of Hawai'i at M\={a}noa, Honolulu, Hawai’i, HI 96822, USA}
\affiliation{$^{9}$Institute for Astrophysics, University of Vienna, 1180 Vienna, Austria}
\affiliation{$^{10}$Department of Physics, Florida State University, Tallahassee, FL 32306, USA}
\affiliation{$^{11}$Carnegie Observatories, Las Campanas Observatory, Casilla 601, La Serena, Chile}
\affiliation{$^{12}$School of Physics, The University of Melbourne, Parkville, VIC, Australia}
\affiliation{$^{13}$Department of Astronomy and Astrophysics, University of California, Santa Cruz, CA, USA}
\affiliation{$^{14}$Department of Physics and Astronomy, Ball State University, Muncie, IN 47306, USA}

\begin{abstract}
We discuss ASASSN-24fw, a 13th-magnitude star that optically faded by $\Delta g = 4.12 \pm 0.02$~mag starting in September 2024 after over a decade of quiescence in ASAS-SN. The dimming lasted $\sim$8 months before returning to quiescence in late May 2025. The spectral energy distribution (SED) before the event is that of a pre-main sequence or a modestly evolved F star with some warm dust emission. The shape of the optical SED during the dim phase is unchanged and the optical and near-infrared spectra are those of an F star. The SED and the dilution of some of the F star infrared absorption features near minimum suggest the presence of a $\sim$0.25$M_\odot$ M dwarf binary companion. The 43.8~year period proposed by \cite{2024ATel16919....1N} appears correct and is probably half the precession period of a circumbinary disk.
The optical eclipse is nearly achromatic, although slightly deeper in bluer filters, $\Delta (g-z)=0.31\pm0.15$ mag, and the $V$ band emission is polarized by up to 4\%. The materials most able to produce such small optical color changes and a high polarization are big ($\sim$20~$\mu$m) carbonaceous or water ice grains. Particle distributions dominated by big grains are seen in protoplanetary disks, Saturn-like ring systems and evolved debris disks. We also carry out a survey of occultation events, finding  46 additional systems, of which only 7 (4) closely match $\varepsilon$ Aurigae (KH 15D), the two archetypes of stars with long and deep eclipses. The full sample is widely distributed in an optical color-magnitude diagram, but roughly half show a mid-IR excess. It is likely many of the others have cooler dust since it seems essential to produce the events.
\end{abstract}

\section{Introduction \label{intro}}

Long-duration and deep stellar dimming events have been serendipitously discovered over the years, and they are traditionally discussed in the context of two archetypes with eclipses caused by disks: $\varepsilon$ Aurigae \citep{1991ApJ...367..278C} and KH 15D \citep{1998AJ....116..261K}. $\varepsilon$ Aur has $\Delta V$=0.75 mag deep, nearly two-year-long eclipses with a period of 27.1 years \citep{2010AJ....139.1254S}. The eclipsed star is a F0 supergiant in a binary system where the companion is a B star surrounded by an optically thick disk \citep{2010ApJ...714..549H}. KH 15D is a pre-main-sequence binary star occulted by a circumbinary disk \citep{2006ApJ...644..510W}. The binary orbital period is 48 days, but the depth and duration of the eclipses are variable due to the precession of the disk. The eclipses can be as deep as 5 mag in the $I$ band \citep{2006ApJ...644..510W,2021MNRAS.503.1599P}.

However, the increasing numbers of long-term time domain surveys have led to the discovery of several dozen systems with long and deep eclipses showing different phenomena. For example, the substructure of the eclipses of 1SWASP~J140747.93$-$394542.6 (V1400 Cen) is explained as occultations by ring structures \citep{2012AJ....143...72M,2014MNRAS.441.2845V,2015MNRAS.446..411K} and the eclipse shape of ASASSN-21js is explained by a ringed disk \citep{2024AA...688L..11P}. Some events are claimed to be created by debris from collisions between orbiting objects, producing irregularly shaped eclipses that are usually aperiodic and of variable duration, examples are KIC 8462852 \citep[commonly known as Tabby's Star,][]{2016MNRAS.457.3988B}, TYC 8830 410 1 \citep{2021ApJ...923...90M}, TIC 400799224 \citep{2021AJ....162..299P}, and ASASSN-21qj \citep{2023ApJ...954..140M,2023Natur.622..251K}.

Long and deep stellar dimming provides a unique laboratory to study stars (or binary stars) surrounded by large, often dusty, structures and to understand the evolutionary and dynamical conditions that can lead to such configurations. Increasing the population of long and deep dimming events will also help to assess whether there are features common to all systems and to systematically categorize them.

The All-Sky Automated Survey for Supernovae (ASAS-SN) project \citep{2014ApJ...788...48S,2017PASP..129j4502K,2023arXiv230403791H}, which consists of 20 telescopes located in Hawai`i, Texas, Chile, and South Africa. ASAS-SN automatically surveyed the entire night sky daily in the $V$ band from 2012 to 2018 and in the Sloan $g$ band since 2018 with a depth of $g\sim18.5$ mag in good conditions. The goal of the project is to identify transient events, primarily supernova explosions, but also tidal disruption events, eclipsing binaries and variable stars.

ASAS-SN has already discovered several long and deep dimming events. For example, ASASSN-V J213939.3$-$702817.4 \citep{2019ATel12836....1J} faded 1.3 magnitudes during 3 days in June 2019 and ASASSN-V J060000.76$-$310027.83 \citep{2019ATel13346....1W} had an irregularly shaped event that started in October 2019 and lasted 580 days with a maximum depth of 0.9 magnitudes. ASASSN-21co \citep{2021ATel14436....1W,2021RNAAS...5..147R} underwent two 0.6~mag dimming events, one in April 2009 and the other in March 2021. The 66-day eclipses can be explained by two M giant stars and the event separation suggests a period of 11.9 years. This has been confirmed with the recent detection of a 0.2~mag secondary eclipse \citep{2025arXiv250719594J}. In June 2021, ASAS-SN reported\footnote{https://www.astronomy.ohio-state.edu/asassn/transients.html} a star undergoing a fading event, ASASSN-21js. The eclipse, which is still ongoing, is asymmetric and as deep as 20\%. \citet{2024AA...688L..11P} modeled it as an eclipse caused by two concentric rings around a substellar object bound to a main-sequence star. The optical dimming event in ASASSN-21qj began on August 2021 \citep{2021ATel14879....1R,2022ATel15531....1R}. In fact, the star started exhibiting an infrared excess 2.5 years before the aperiodic, irregular optical obscuration began and lasted for 500 days. \citet{2023Natur.622..251K} and \citet{2023ApJ...954..140M} both explain the event with a collision creating a debris cloud that later transits in front of the star. ASASSN-24fa \citep{2024ATel16715....1J} is a short-timescale variable that has been undergoing a 0.3~mag dimming event since March 2022. Recently, ASAS-SN reported two 0.9~mag dimmings separated by a month in ASASSN-25bv \citep{2025ATel17196....1J}.

\begin{figure*}
\centering
\includegraphics[width=\linewidth]{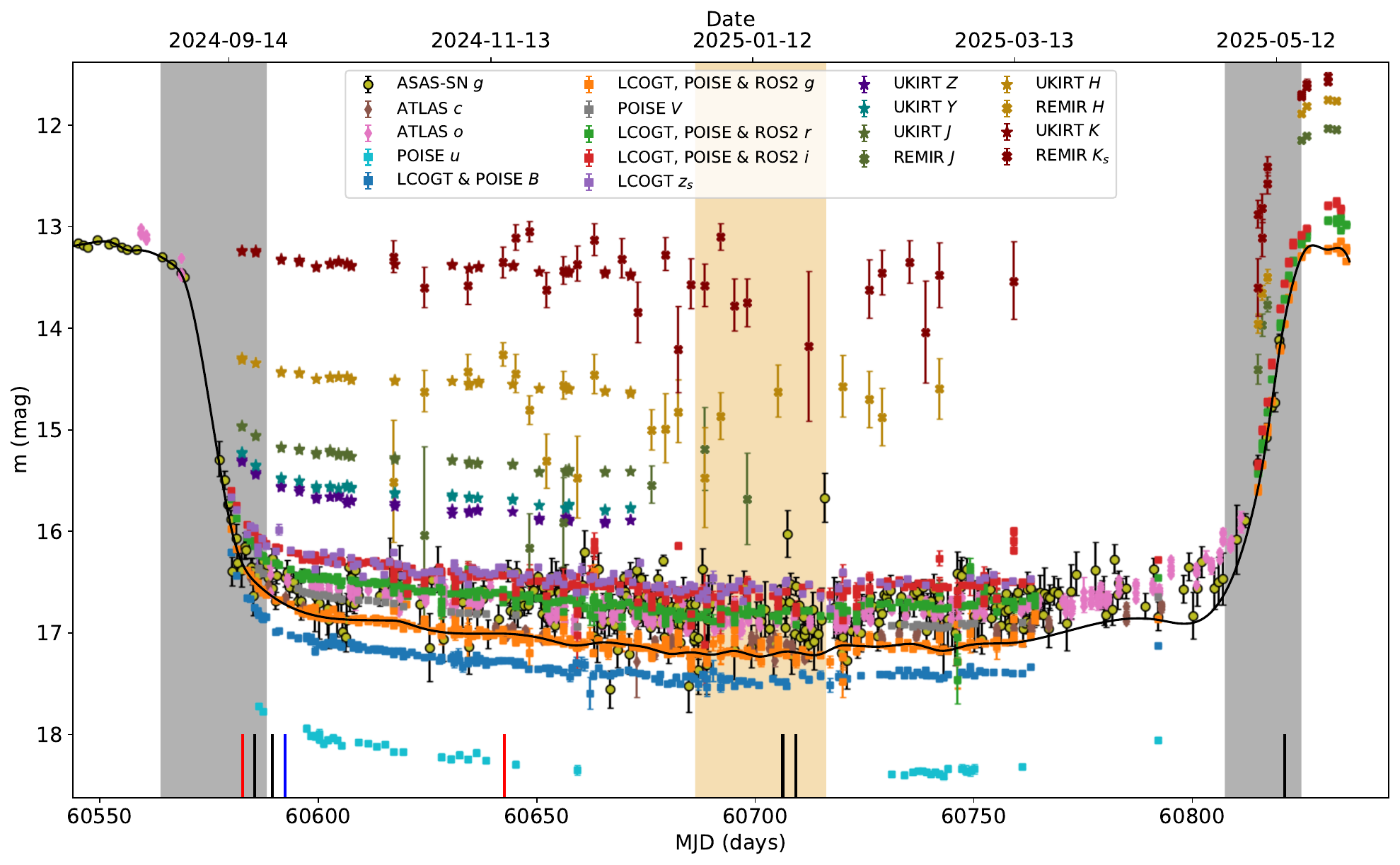}
\caption{The light curve of the event. The grey shaded band shows
the ingress (egress) period of 24 (18) days over which the magnitude of the star drops from 0.2 to 3.5 mag with respect to the quiescent magnitude. The LCOGT, POISE and ROS2 $u$, $B$, $g$, $V$, $r$, $i$ and $z_s$ light curves are shown with squares and the ATLAS $c$ and $o$ light curves are shown with brown and pink diamonds. The UKIRT $Z$, $Y$, $J$, $H$ and $K$ light curves are shown as stars and the REMIR $J$, $H$ and $K_s$ light curves are shown as crosses. The light curve minimum is shadowed in light orange. The solid line is a spline model of the ASASSN/LCOGT/POISE/ROS2 g-band light curve. It also very closely matches the data if reversed around MJD 60697. The black, blue and red vertical lines at the bottom mark when polarization measurements, the optical spectrum and the infrared spectra, respectively, were taken. \label{fig:lc_in}}
\end{figure*}

On September 27, 2024, ASAS-SN reported another deep dimming event, ASASSN-24fw \citep{2024ATel16833....1J}. Previously, the star showed no variability since first observed by ASAS-SN in February 2012. It then developed a deep (4.1 mag), symmetric, obscuring event lasting 8 months before recovering at the end of May 2025, as shown in Fig.~\ref{fig:lc_in}. In this paper, we study this nearly total eclipse and discuss scenarios for its origin. In Section \ref{sec:obs} we describe the data used to characterize this system and the properties of the quiescent system. Section \ref{sec:interp} is devoted to the analysis of the observations and inferring the properties of the occulted star, the occulter, and the orbit and geometry of the system. We put this event into context with a survey of long and deep eclipses in Section \ref{sec:pop} and summarize the results in Section \ref{sec:concl}. A simultaneous article about ASASSN-24fw has been presented independently by \citet{2025arXiv250705367Z} with complementary observations during the occultation.

\section{Observations}\label{sec:obs}

The quiescent star associated with ASASSN-24fw was observed by several different surveys before the dimming event. Table \ref{tab:24fw_astro} summarizes the {\it Gaia} \citep{2016A&A...595A...1G,2023A&A...674A...1G} astrometric properties, the photogeometric distance from \citet{2021AJ....161..147B} and the estimated foreground extinction based on the \texttt{mwdust} \citep{2016ApJ...818..130B} 3D dust models (a combination of \citealt{2003A&A...409..205D}, \citealt{2006A&A...453..635M}, and \citealt{2019ApJ...887...93G}).

\begin{table}
\caption{{\it Gaia} DR3 astrometric solutions, \citet{2021AJ....161..147B} distance and \texttt{mwdust} pre-event foreground extinction of ASASSN-24fw.}
\label{tab:24fw_astro}
\centering
\begin{tabular}{l c}
\hline\hline
Parameter & Value \\
\hline
    Target ID & 3152916838954800512 \\
    RA$_{J2016}$ & 106.3291 deg \\
    DEC$_{J2016}$ & 6.2054 deg \\
    Galactic l$_{J2016}$ & 208.9677 deg \\
    Galactic b$_{J2016}$ & 5.9024 deg \\
    Parallax & 0.9583$\pm$0.0152 mas \\
    RA proper motion & $-$3.7500$\pm$0.0162 mas yr$^{-1}$ \\
    DEC proper motion & $-$7.6121$\pm$0.0144 mas yr$^{-1}$ \\
    Distance & $1011^{+15}_{-23}$ pc \\
    $E(B-V)$ & 0.062 mag \\
\hline
\end{tabular}
\end{table}

\subsection{Archival photometry}\label{subsec:arcphot}

Table \ref{tab:phot_prev} summarizes the available archival photometry of the star. We include $u/v$ data from SkyMapper \citep[DR4 doi:10.25914/5M47-S621,][]{2024PASA...41...61O}, $g/z$ data from the Panoramic Survey Telescope and Rapid Response System \citep[Pan-STARRS,][]{2016arXiv161205560C,2020ApJS..251....7F}, $g/r/i/z$ data from the ATLAS All-Sky Stellar Reference Catalog \citep[REFCAT, doi:10.17909/t9-2p3r-7651,][]{2018ApJ...867..105T}, $J/H/K_s$ data from the Two Micron All-Sky Survey \citep[2MASS,][]{2003tmc..book.....C} and $W1/W2/W3/W4$ (3.4/4.6/12/22$\mu$m, respectively) data from AllWISE \citep{2010AJ....140.1868W,2011ApJ...731...53M}. The magnitudes are reported in their natural photometric system, (i.e., the Vega system for 2MASS and WISE and the AB system for the rest).

\begin{table}
\caption{Photometric magnitudes used to fit the SED of ASASSN-24fw before the dimming.}
\label{tab:phot_prev}
\centering
\begin{tabular}{c c r@{$\pm$}l}
\hline\hline
 Source & Band & \multicolumn{2}{c}{Mag} \\
\hline
\multirow{2}{*}{SkyMapper} & $u$ & 14.30&0.06\\
                           & $v$ & 13.85&0.04 \\
\hline
\multirow{2}{*}{Pan-STARRS} & $g$ & 13.1212&0.0010 \\
                             & $z$ & 12.796&0.003 \\
\hline
\multirow{4}{*}{REFCAT} & $g$ & \multicolumn{2}{c}{13.08} \\
                          & $r$ & \multicolumn{2}{c}{12.85} \\
                          & $i$ & \multicolumn{2}{c}{12.79} \\
                          & $z$ & \multicolumn{2}{c}{12.79} \\
\hline
\multirow{3}{*}{2MASS} & $J$ & 12.01&0.02 \\
                          & $H$ & 11.76&0.02 \\
                          & $K_s$ & 11.53&0.02 \\
\hline
\multirow{4}{*}{AllWISE} & $W1$ & 10.76&0.02 \\
                          & $W2$ & 9.97&0.02 \\
                          & $W3$ & 7.191&0.015 \\
                          & $W4$ & 5.67&0.04 \\
\hline
\end{tabular}
\end{table}

\subsection{Current photometry}

\begin{table}
\caption{Photometric magnitudes of ASASSN-24fw during the dimmest phase (from January 11 to February 10, 2025).}
\label{tab:phot_post}
\centering
\begin{tabular}{c c c}
\hline\hline
 Source & Band & Mag \\
\hline
\multirow{5}{*}{LCOGT} & $B$ & 17.48$\pm$0.03 \\
 & $g'$ & 17.20$\pm$0.02 \\
 & $r'$ & 16.80$\pm$0.02 \\
 & $i'$ & 16.59$\pm$0.04 \\
 & $z_s$ & 16.55$\pm$0.05 \\
\hline
\multirow{2}{*}{POISE} & $u$ &  18.41$\pm$0.04 \\
 & $V$ & 16.99$\pm$0.03 \\
\hline
\multirow{2}{*}{ATLAS} & $c$ &  17.14$\pm$0.06 \\
 & $o$ & 16.87$\pm$0.06 \\
\hline
\multirow{5}{*}{UKIRT} & $Z$ & 15.89$\pm$0.03 \\
 & $Y$ & 15.75$\pm$0.04 \\
 & $J$ & 15.40$\pm$0.03 \\
 & $H$ & 14.60$\pm$0.03 \\
 & $K$ & 13.45$\pm$0.03 \\
\hline
\end{tabular}
\end{table}

The ASAS-SN project has monitored the star since February 16, 2012 with 4346 epochs. The star was observed in the $V$ band until November 28, 2018 with a stable magnitude of $V=12.90$ mag and a standard deviation about the mean of 0.02 mag. It was observed in the $g$ band from April 12, 2018 until the start of the dimming event, with a mean magnitude of 13.11 mag and a dispersion of 0.03 mag. The average magnitude during the dim phase is $g=16.5\pm0.6$ mag, so the star decreased in flux by 95.5\%. The light curve during the dimming is shown in Figure \ref{fig:lc_in}.

Las Cumbres Observatory Global Telescope \citep[LCOGT, ][]{2013PASP..125.1031B} started to follow the star with the 1~m telescope network in the $B/g'/r'/i'/z_s$ filters after the beginning of the dimming on September 26, 2024 until March 28, 2025 with over 130 measurements in each band. The light curves are shown in Figure \ref{fig:lc_in} where the data acquired during full moon have a larger scatter, particularly in the $B$ band. All light curves show a slight decrease until January 2025 and then a steady increase. Table \ref{tab:phot_post} shows the magnitudes and standard deviation of the faintest phase (light orange shaded region in Figure \ref{fig:lc_in}).

The Precision Observations of Infant Supernova Explosions (POISE) collaboration \citep{2021ATel14441....1B} used the 1~m Swope Telescope at Las Campanas Observatory to follow the star from October 3, 2024 until April 27, 2025 in $u/B/V/g/r/i$ bands with around 45 measurements. The photometry is in the CSPII natural system \citep{2019PASP..131a4001P} and we transform the LCOGT photometry to this system using the color terms from \citet{Burns_inprep}. Due to the gap in POISE measurements during the minimum brightness phase (see Figure \ref{fig:lc_in}), interpolated $u$ and $V$ magnitudes in this phase are presented in Table \ref{tab:phot_post}.

We used the 3.8~m United Kingdom Infrared Telescope (UKIRT), on Maunakea, Hawai`i, to acquire 38 epochs of near-infrared (NIR) $Z/Y/J/H/K$ measurements from September 29 to December 27, 2024. Since the measurements do not cover the light curve minimum (see Figure \ref{fig:lc_in}), we average of the measurements for the last 30 days in Table \ref{tab:phot_post}.

We monitored ASASSN-24fw with the Rapid Eye Mount (REM) telescope at La Silla, Chile between November 3, 2024 and June 8, 2025 (director's discretionary time, E. Gaidos, PI).  REM is a 60-cm automated telescope with separate optical (ROS2) and near-infrared (REMIR) detectors obtaining $10\arcmin \times 10\arcmin$ images in the Sloan $g/r/i/z$ and 2MASS-like $J/H/K_s$ passbands, respectively \citep{2001AN....322..275Z,2019lsof.confE..18M}. We obtained 51 epochs of data with ROS2 and 36 with REMIR, with per visit integration times of 180~s in $g/r/i/z$ and 75 or 150~s in $J/H/K_s$.   

\subsection{Historical and other surveys}

\begin{figure*}
\centering
\includegraphics[width=0.95\linewidth]{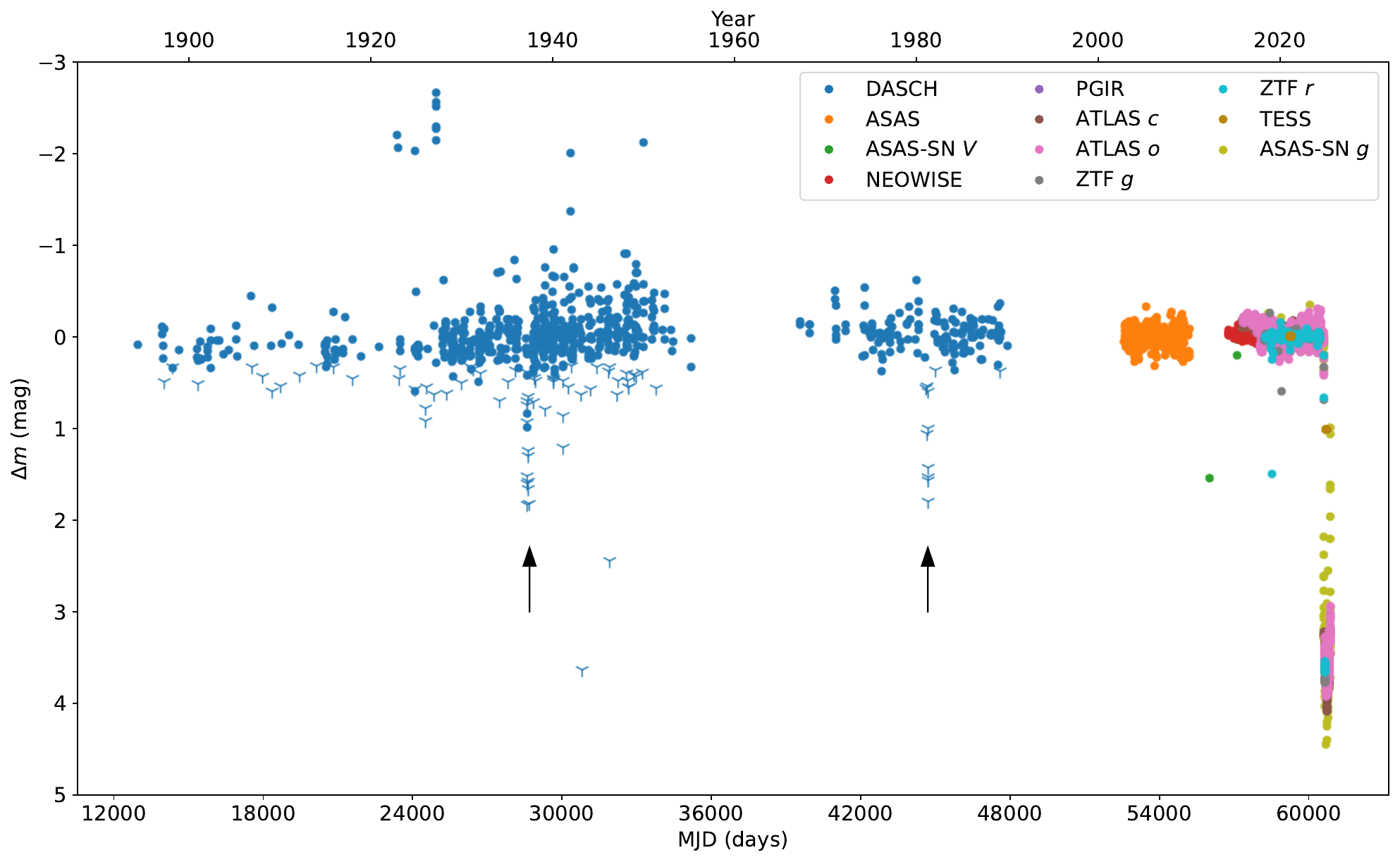}
\includegraphics[width=0.95\linewidth]{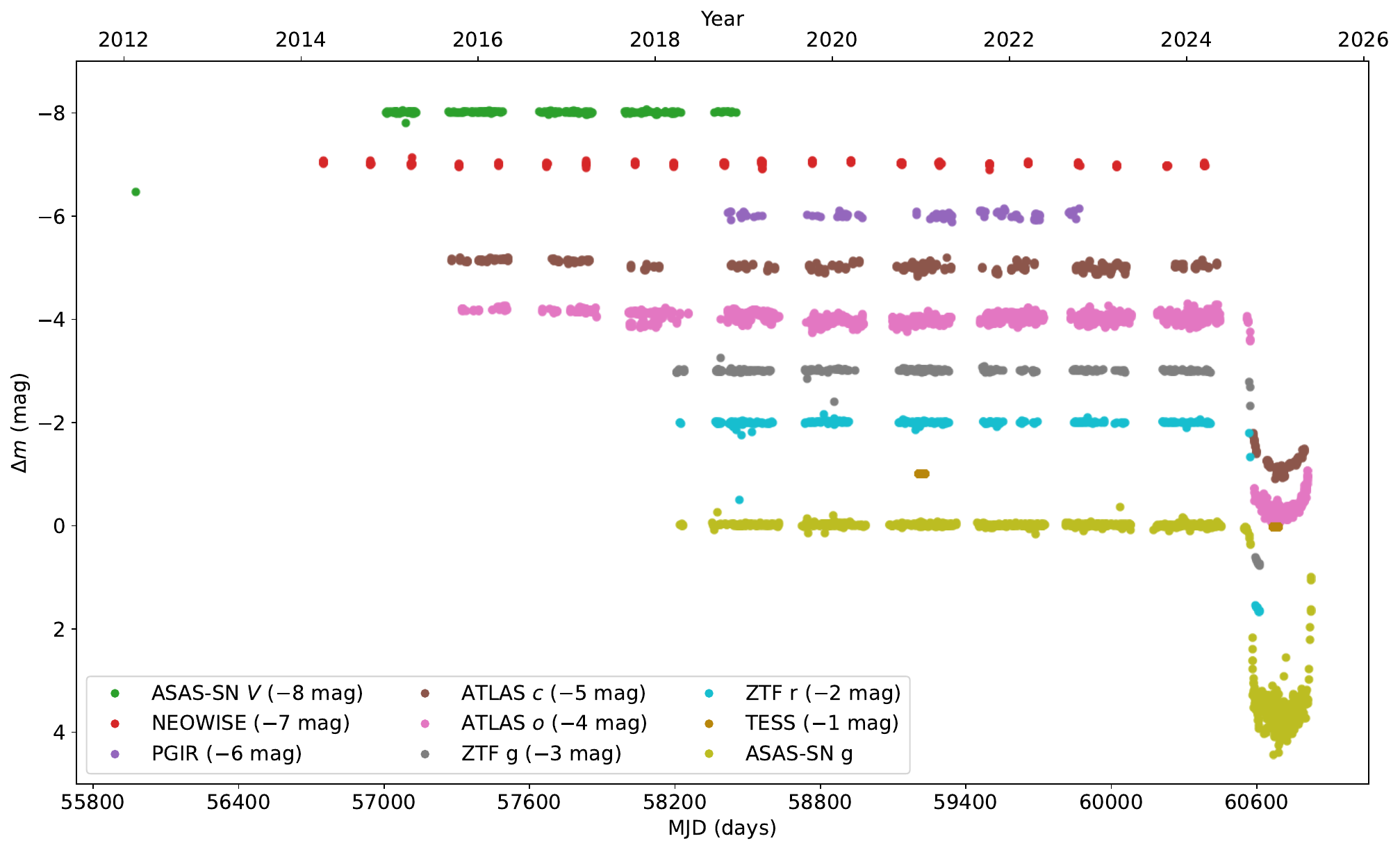}
\caption{The top panel shows all the data from 1894 until the present time normalized to the median magnitude of each time series. The down facing blue symbols are the DASCH magnitude limits fainter than the standard deviation of the DASCH time series. The black arrows mark the possible past events reported by \citet{2024ATel16919....1N}. The bottom panel shows the recent light curves shifted in magnitude to better show them individually. \label{fig:all_lcs}}
\end{figure*}

ASASSN-24fw has been monitored by several other surveys over the years. The Digital Access to a Sky Century @ Harvard project \citep[DASCH, see][]{2010AJ....140.1062L,2012IAUS..285...29G,2013PASP..125..857T} has 95 years of photographic plate observations of the star from 1894 to 1989 with 687 measurements in the photographic $B$ band. The All-Sky Automated Survey \citep[ASAS,][]{1997AcA....47..467P} monitored the star from October 2002 to November 2009 in the $V$ band with 389 high quality measurements. The Near-Earth Object Wide-field Infrared Survey Explorer (NEOWISE) mission \citep{2011ApJ...731...53M} has $W1$ and $W2$ light curves covering from April 2014 to March 2024 with 286 measurements. Unfortunately, these measurements stopped just before the dimmimg. The Palomar Gattini-IR (PGIR) survey \citep{2024PASP..136j4501M} has 107 measurements of the source from October 2018 to October 2022 in the $J$ band. The star also has photometry from the Zwicky Transient Facility \citep[ZTF, Data Release 23, ][]{2019PASP..131a8003M,2019PASP..131a8002B} in the $g$ and $r$ bands with 353 and 504 epochs, respectively, from March 2018 to October 2024. The Asteroid Terrestrial-impact Last Alert System (ATLAS) project \citep{2018ApJ...867..105T,2020PASP..132h5002S} has photometry of the star from September 12, 2015 until May 15, 2025. We used the ATLAS forced photometry server \citep{2021TNSAN...7....1S} to extract 705 and 2621 measurements in the $c$ and $o$ bands, respectively. 

Lastly, the Transiting Exoplanet Survey Satellite \citep[TESS,][]{2015JATIS...1a4003R} observed the star in Sector 33 which spans from December 18, 2020 to January 13, 2021 with 3461 observations and in Sector 87 (from December 20, 2024 to January 13, 2025) with 9360 observations. The TESS light curves were extracted and detrended using the {\tt unpopular} package \citep{2022AJ....163..284H}. Although TESS observed both the quiescent and obscured phases, the systematics of the different sectors prevent us from properly determining the depth of the eclipse in the TESS band. We find a weak periodic signal at 9.58 days in the pre-dimming TESS light curve. However, after considering contamination using \texttt{TESS-Localize} \citep{Higgins2023}, we conclude that the signal is likely due to contamination from nearby stars. No periodic signal is found for Sector 87.

\subsection{Spectroscopy}

\begin{figure*}
\centering
\includegraphics[width=\linewidth]{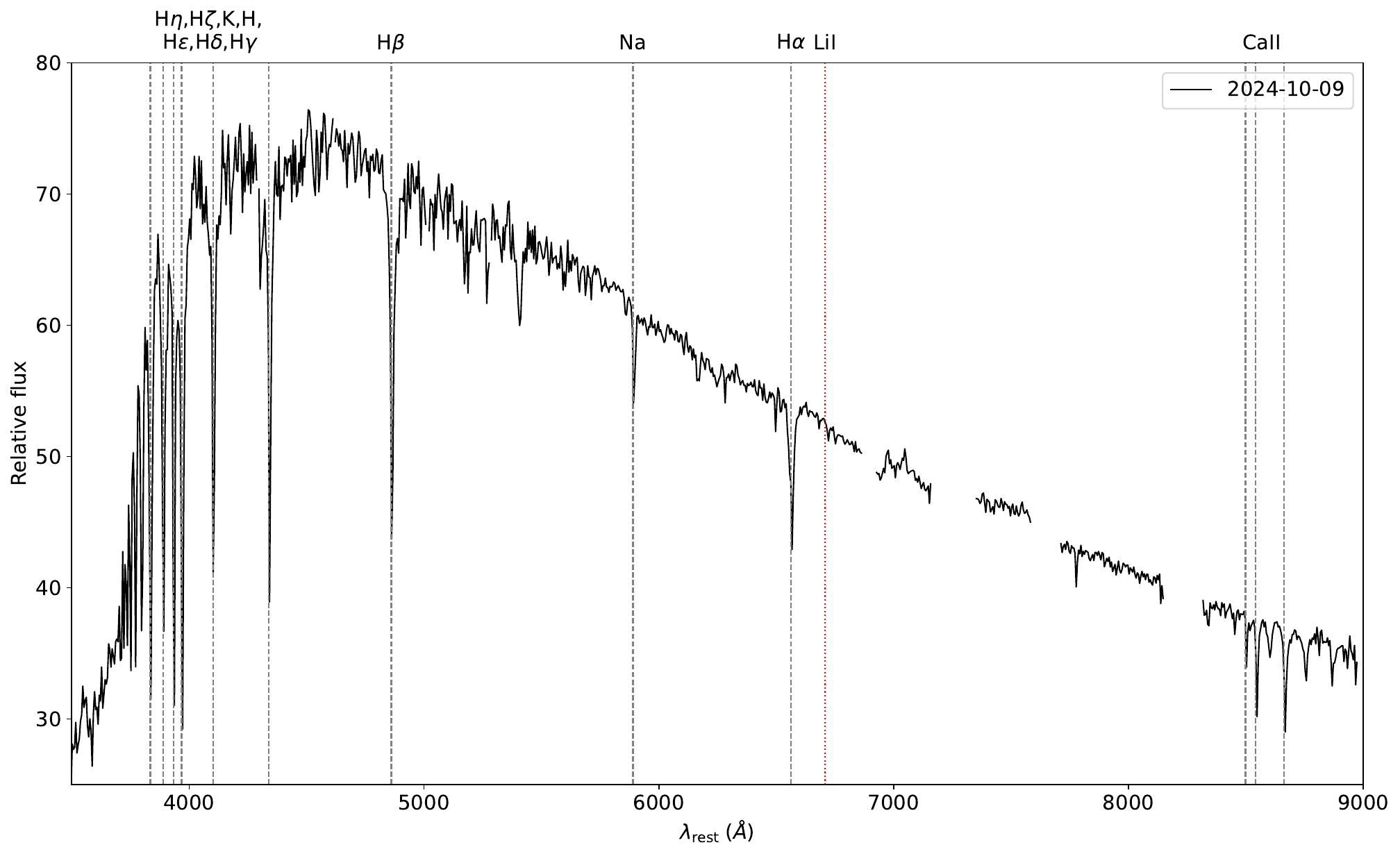}
\includegraphics[width=\linewidth]{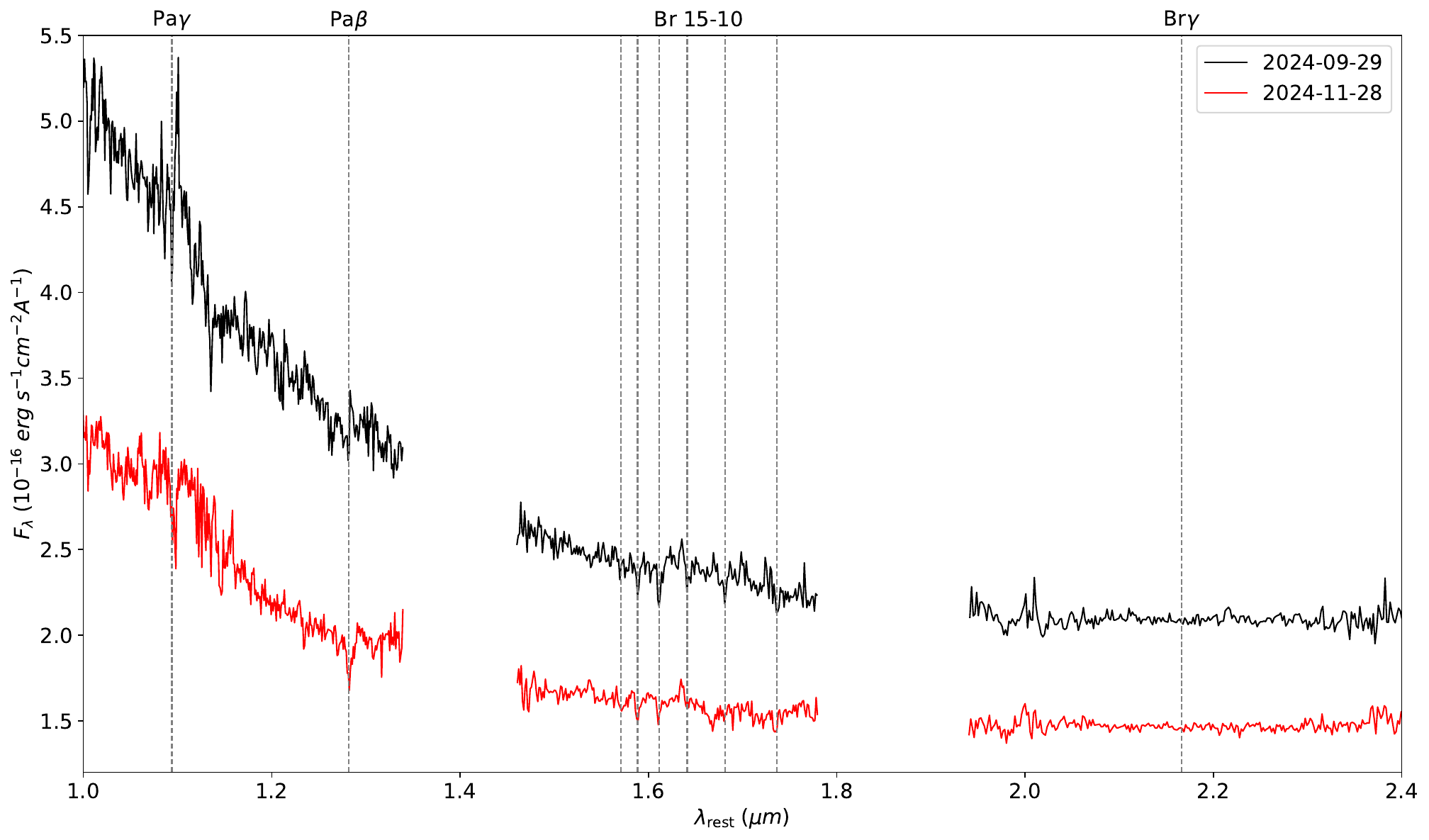}
\caption{LBT MODS optical (top) and IRTF SpeX near-IR (bottom) spectra acquired during the dim phase. The most common absorption lines found in F stars are marked with dashed vertical lines along the wavelength of any Li absorption at 6708 \AA. \label{fig:spec}}
\end{figure*}

We could find no pre-event spectra, but we acquired two long-slit spectra from the 3.0 m NASA Infrared Telescope Facility (IRTF) on Maunakea with the SpeX spectrograph \citep{2003PASP..115..362R} on September 29 and November 28, 2024. The total integration times were 28 and 52 minutes, respectively, with a slit width of 0\farcs8 and covering a wavelength range from 0.8 to 2.4 $\mu$m at a spectral resolution of R$\sim$2500 in the SXD mode. Reductions were performed using the Spextool IDL package \citep{2004PASP..116..362C} following standard procedures and an A0 star was used for telluric corrections. We also obtained an optical spectrum on October 9, 2024 using the Multi-Object Double CCD Spectrographs \citep[MODS,][]{2010SPIE.7735E..0AP} mounted on the twin 8.4-meter diameter Large Binocular Telescope (LBT). The exposure time was 15 minutes with a spectral range from 3100 to 10500 \AA\ and a resolution of R$\sim$2000.  The spectra are shown in Figure \ref{fig:spec}.

\subsection{Polarization}

We obtained polarimetric and photometric observations of ASASSN-24fw at the 1.83~m Perkins telescope (Flagstaff, AZ) with the PRISM camera\footnote{https://www.bu.edu/prism/} equipped with a rotating polaroid  (POL-HN38) and a wheel of standard UBVRI filters. The observations were performed on October 2 and 6, 2024, and January 31, February 3 and May 26, 2025 in $V$ band at an effective wavelength $\lambda_{\text {eff}}$=551~nm. These involved a series of 4 Stokes I, Q and U measurements. Each series consists of four measurements at polaroid position angles (P.A.) of 0$^\circ$, 45$^\circ$, 90$^\circ$, and 135$^\circ$, with an exposure time of 150~s for each measurement. The four measurements of the I, Q, and U parameters were averaged to calculate the degree of polarization, $P=\sqrt{Q^2+U^2}$, and its position angle, $\chi=0.5\arctan{(U/Q)}$, and their uncertainties. Since the camera has a wide $14\arcmin\times14\arcmin$ field of view, we used field stars to perform both interstellar and instrumental polarization corrections, assuming that stars in the field are intrinsically unpolarized. We used unpolarized calibration stars from \citet{1992AJ....104.1563S} to check the instrumental polarization, which is usually less than 0.2\%, and polarized stars from the same paper to calibrate the polarization P.A.

Table \ref{tab:pol} shows the polarization measurements. We find a slight correlation between $V$ magnitude and polarization (Pearson correlation coefficient $r=0.54^{+0.18}_{-0.21}$) and a strong anti-correlation between magnitude and polarization P.A. ($r=-0.90^{+0.09}_{-0.06}$). However, no correlation is found between these two parameters if we drop the last measurement ($r=0.02^{+0.42}_{-0.62}$). The 68\% confidence intervals are computed using Monte Carlo methods to account for the uncertainties in the measurements.

\begin{table}
    \caption{Polarization measurements of ASASSN-24fw during the dimming event. The columns are the Modified Julian Date of observation, $V$ magnitude, linear polarization fraction, and polarization position angle.}
    \label{tab:pol}
    \centering
    \begin{tabular}{cr@{$\pm$}lr@{$\pm$}lr@{$\pm$}l}
    \hline\hline
MJD & \multicolumn{2}{c}{$V$ (mag)} & \multicolumn{2}{c}{P (\%)} & \multicolumn{2}{c}{$\chi$ (deg)} \\
\hline
60585.4668 & 16.258&0.027 & 2.93&0.58 & 107.9&5.9 \\
60589.4453 & 16.198&0.024 & 1.78&0.50 &  93.9&10.5 \\
60706.2822 & 16.939&0.042 & 3.97&0.60 &  97.1&4.3 \\
60709.2402 & 16.854&0.031 & 2.80&0.57 & 105.1&5.9 \\
60821.1445 & 13.773&0.007 & 2.01&0.40 & 143.2&5.6\\
\hline
    \end{tabular}
\end{table}

\begin{figure}
\centering
\includegraphics[width=\linewidth]{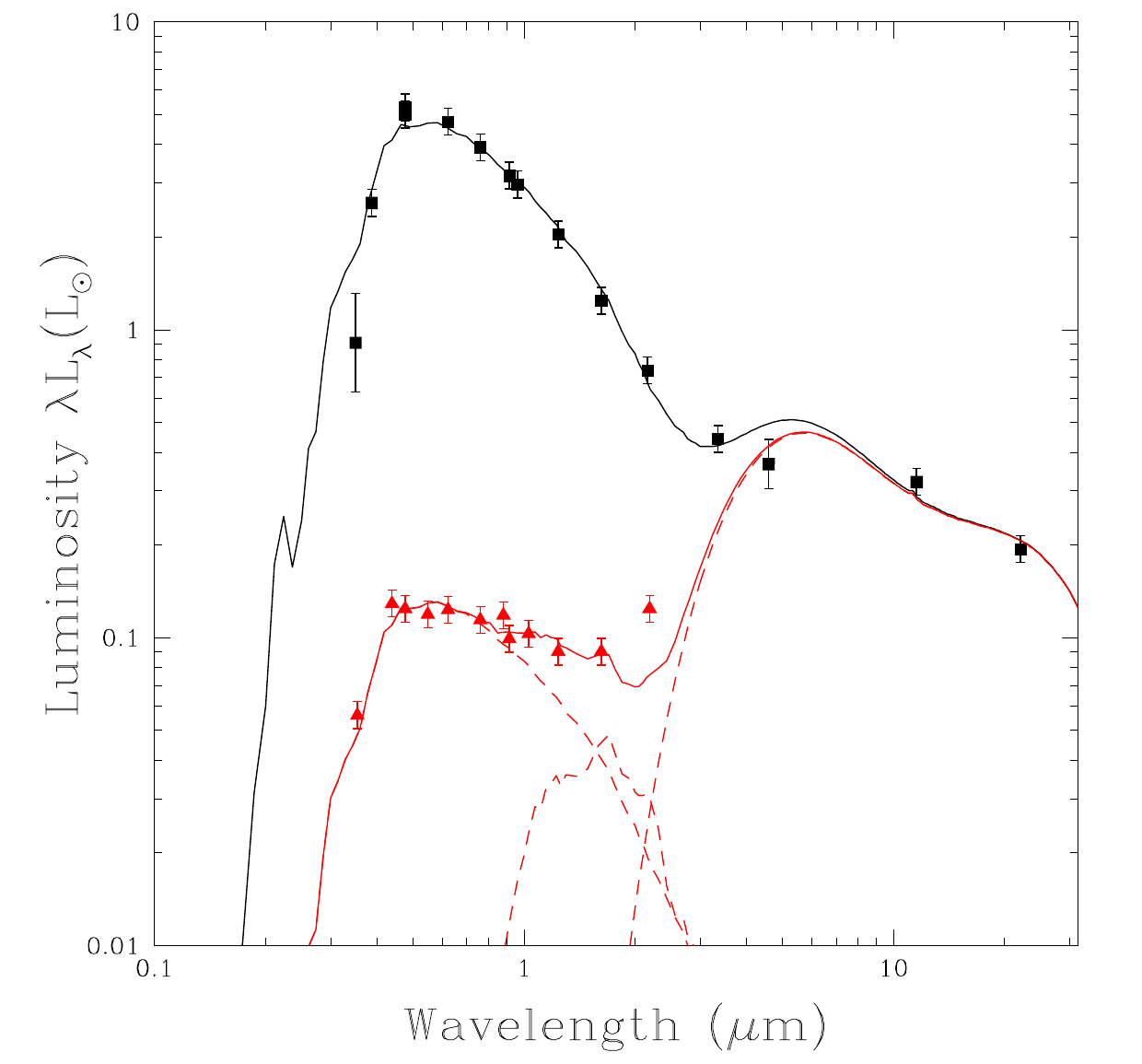}
\caption{The foreground extinction-corrected SED of the quiescent star (black squares) and the model fit (black solid line).  The red triangles are the SED at the light curve minimum. The red solid curve is an approximate model for the SED combining the three
components shown by the dashed red curves.  The component on the left is the non-dust emission from the star from the SED model, simply reducing the luminosity of the non-dust emission of the model to match the optical fluxes. The middle component is the SED of an $0.25 M_\odot$ M dwarf companion.  The right
 component is the dust emission associated with the
 quiescent star.
 \label{fig:SED}}
\end{figure}

\begin{figure}
\centering
\includegraphics[width=\linewidth]{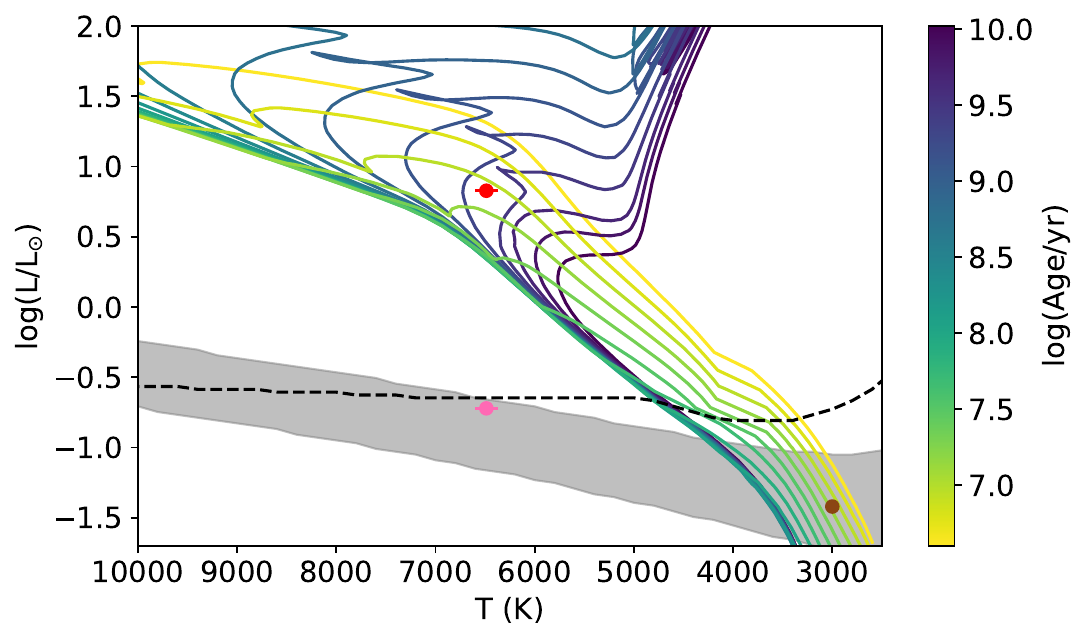}
\caption{Inferred luminosity and temperature from fitting the quiescent (red) and obscured (pink) SEDs. Solar metallicity isochrones colored by age are shown as well as the luminosity limits for a companion black body that does not significantly change the SED shape of the dim phase (dashed black line) and dilutes Br$\gamma$ IR line but does not completely mask the the Pa$\beta$ line (grey band). The point in the lower right corner is the proposed M dwarf companion from Figure~\ref{fig:SED}. \label{fig:LT_iso}}
\end{figure}

\section{Data interpretation}\label{sec:interp}

We discuss five aspects of the event. First, we characterize the unobscured star in Section \ref{subsec:star}. Then we quantify the properties of the dimmed star in Section \ref{subsec:ecl}. Next, we discuss the possible orbital parameters in Section \ref{subsec:orb}. The physical properties and the origin of the obscuring material are discussed in Section \ref{subsec:occ}. Finally, we discuss possible system geometries in Section~\ref{subsec:geom}.

\subsection{The Quiescent Star}\label{subsec:star}

We modeled the pre-event spectral energy distribution (SED) using the archival photometry in Table~\ref{tab:phot_prev}, the reported central value of the distance and a foreground extinction of $E(B-V)=0.062$ mag.  We fit the foreground extinction-corrected SED as in \citet{2017MNRAS.468.4968A}, running {\tt DUSTY} \citep{2001MNRAS.327..403E} inside a Markov Chain Monte Carlo (MCMC) driver to both optimize the fits and then estimate the uncertainties.  {\tt DUSTY} solves the dust radiation transport problem for a central star surrounded by a spherical shell of dust. We used \citet{2003IAUS..210P.A20C} or MARCS (\citealt{Gustafsson2008}) model stellar atmospheres.  We varied the stellar luminosity $L_*$, effective temperature $T_*$, dust temperature $T_d$ and dust visual optical depth $\tau_V$. We used a circumstellar dust extending from an inner radius $R_{in}$ from the star to an outer radius $R_{out}=4R_{in}$ where the inner radius is determined by the other parameters of the model and the ratio $R_{out}/R_{in}=4$ is fixed because the model is unable to constrain it without longer wavelength measurements. We used the \citet{1984ApJ...285...89D} graphitic dust model with an \citet{1977ApJ...217..425M} size distribution ($dn/da \propto a^{-3.5}$ with $0.005\mu\hbox{m} \leq a \leq 0.25\mu\hbox{m}$). Silicate dust models with this size distribution poorly fit the SED because they have a strong absorption feature at $\sim$10~$\mu$m. We use minimum errors of 10\% for the SED, and artificially increased the uncertainties on the Skymapper $u$ band (to allow a better fit of the peak of the SED) and the WISE
$4.5\mu$m band (to better fit the $3.6\mu$m band and allow a solution with somewhat hotter dust, see below).

The model fits the data (see Figure \ref{fig:SED}) with a reduced $\chi^2$ of 1.12 for 11 degrees of freedom and an effective temperature of $T_{\ast}=6490\pm120$ K, a luminosity of $\log(L_{\ast}/L_{\odot})=0.828\pm0.005$ and a radius of $R_{\ast}=2.05\pm0.07\; R_{\odot}$. The inferred dust visual optical depth is $\tau_{V}=0.249\pm0.017$, with a temperature of $T_{d}=552\pm16$ K and an inner radius of $\log(R_{in}/\text{cm})=13.85\pm0.03$. These results are for a fixed distance of 1011 pc, and for the upper (lower) limits on the distance, the luminosity logarithm, stellar radius and dust inner radius logarithm would increase by 1.5, 1.5, and 0.05\% (decrease by 2.4, 2.3, and 0.07\%), respectively. Based on the estimate of $T_{\ast}$, the quiescent source is an F star.

By matching the inferred $L_{\ast}$ and $T_{\ast}$ 2$\sigma$ intervals to Solar metallicity {\tt PARSEC} \citep{2012MNRAS.427..127B,2014MNRAS.444.2525C,2015MNRAS.452.1068C,2014MNRAS.445.4287T,2017ApJ...835...77M,2019MNRAS.485.5666P,2020MNRAS.498.3283P} isochrones (see Figure \ref{fig:LT_iso}), we find two possible solutions: a young star with a median $\log(\text{Age}/yr)=$7.02 (and a range from 6.94 to 7.08), or an older star with a $\log(\text{Age}/yr)$ of 9.35 (9.31-9.53). The median stellar mass for the younger solution is 1.53$M_{\odot}$ (1.48-1.60$M_{\odot}$) while the older solution is slightly less massive, 1.45$M_{\odot}$ (1.34-1.46$M_{\odot}$). In the younger, higher mass solution the star is still a pre-main sequence star, while in the older, lower mass solution, it has just started evolving off the main sequence. The IR excess in the SED may be due to a protoplanetary disk if the star is young. Since there is no unique age solution, we use the possible mass range of $M_{\ast}=1.45^{+0.15}_{-0.11}M_{\odot}$.

While the spectra in Figure \ref{fig:spec} were taken during the occultation, they show many typical F star features (e.g., H and Ca absorption features), as marked with the dashed lines. Interestingly, in the IR spectra, the Brackett 10--15 lines are clearly detected but Br$\gamma$ line is not, and in the later spectrum these lines are even shallower. Usually, Br$\gamma$ is more prominent than Br 10--15 \citep{2009ssc..book.....G} which may imply that a continuum emission source is diluting the lines in the redder parts of the IR spectra. The spectra are those of an F-type star, consistent with both the quiescent and occulted SEDs.  The presence of lithium in the optical spectrum would favor the young age solution of the SED fit. For a 10 Myr-old F star with $T_\ast\sim6500$ K, we would expect a Lithium 6708 \AA\ line with an equivalent width of $\sim $75 m\AA\ \citep{2023MNRAS.523..802J}.  Given the resolution of MODS and the signal-noise-ratio of the spectrum, we cannot rule out the existence of an absorption feature with an equivalent width less than $\sim$100 m\AA.

\subsection{Photometric properties of the eclipse}\label{subsec:ecl}

The depth of the event can be described by the optical depth, $\tau$, of the occulter and the fraction of the star that it does not cover, $f$,
\begin{equation}
   \Delta m=-2.5\log\left[f+(1-f)e^{-\tau}\right] .
\end{equation}
This relation has two limiting cases: if we assume that the star is fully covered ($f=0$), we can determine the minimum optical depth, $\tau_{\text{min}}$, needed to produce the dimming or, if the obscuring material is completely opaque ($\tau\rightarrow\infty$), we can determine the maximum fraction of the star not covered by the occulter, $f_{\text{max}}$. Table \ref{tab:depth} shows the depth at the minimum in the $g/r/i/z$ bands and the implied limits on the optical depth and uncovered fraction at the time of minimum brightness. We use the mean LCOGT magnitudes near minimum (the light orange band in Figure \ref{fig:lc_in}) compared to the pre-event REFCAT magnitudes. The drop in ASAS-SN, ROS2 and ATLAS are consistent but much less precise, and the ZTF and POISE data do not cover the time of minimum. The optical drop in flux is approximately a factor of 40, requiring an optical depth of $\tau_{\text{min}} \simeq 3.6$ or a maximum uncovered fraction of $f_{\text{max}} \simeq 2.5\%$. While there is a slight chromaticity, with slightly larger drops in the bluer bands, the event is remarkably achromatic given the amplitude. The $g-r$ color at minimum differs from the REFCAT color by only 0.17$\pm$0.03 mag. The color changes during the whole event from LCOGT and POISE measurements are nearly constant and also small with $\Delta( g- r)=0.16\pm0.02$, $\Delta (g-i)=0.29\pm0.03$ and $\Delta( g-z)=0.31\pm0.15$ mags.

\begin{table}
    \caption{Magnitude depth, $\Delta m$, of the dimming in different bands, the minimum optical depth, $\tau_{\text{min}}$, needed to produce the dimming and maximum percentage, $f_{\text{max}}$, of the uncovered area of the star assuming an opaque occulter with no scattering.}
    \label{tab:depth}
    \centering
    \begin{tabular}{c r@{$\pm$}l r@{$\pm$}l r@{$\pm$}l}
    \hline\hline
Band & \multicolumn{2}{c}{$\Delta m$ (mag)} & \multicolumn{2}{c}{$\tau_{\text{min}}$} & \multicolumn{2}{c}{$f_{\text{max}}$ (\%)} \\
\hline
$g$ & 4.12&0.02 & 3.80&0.02 &  2.24&0.04 \\
$r$ & 3.95&0.02 & 3.64&0.02 & 2.64&0.06 \\
$i$ & 3.80&0.04 & 3.50&0.03 & 3.01&0.10 \\
$z$ & 3.76&0.05 & 3.47&0.05 & 3.13&0.15 \\
$J$ & 3.39&0.04 & \multicolumn{2}{c}{\nodata} & \multicolumn{2}{c}{\nodata} \\
$H$ & 2.84&0.03 & \multicolumn{2}{c}{\nodata} & \multicolumn{2}{c}{\nodata} \\
$K_s$ & 1.9&0.4 & \multicolumn{2}{c}{\nodata} & \multicolumn{2}{c}{\nodata} \\
\hline
    \end{tabular}
\end{table}

Table \ref{tab:phot_post} and Figure \ref{fig:SED} show the SED of the source at minimum.  If we consider only the optical data, the shape is essentially 
identical to the SED in quiescence.  Thus, as one component of a model, we use the non-dust emission of the quiescent source model reduced
by an achromatic factor of $36$ to $0.19L_\odot$.  Not surprisingly, the occulted source lies far below any stellar isochrone (see Figure \ref{fig:LT_iso}).  This means that the optical emission at minimum cannot be exclusively from a companion unless it is similarly occulted.

The shape of the SED at longer wavelengths is clearly  different.  We also see this in Table~\ref{tab:depth}, where the depth at minimum in the near-IR (2MASS $J/H/K_s$ versus UKIRT $J/H$ and REMIR $K_s$) is significantly less than in the optical.  Attempts to fit this SED using DUSTY failed.  Given the estimated visual optical depth of the occulter, it would be unable to obscure the mid-IR emission seen in the SED of the quiescent star even if it was large enough to do so, so a second emission component at the minimum is likely the dust emission of the quiescent {\tt DUSTY} SED model.  This plausibly explains the upturn of the SED at $K$ band.  While we increased the uncertainties on the WISE $4.5\mu$m flux in the pre-event photometry to force the SED model to pass through the WISE $3.6\mu$m point and have hotter dust (see Section \ref{subsec:star}), the model still under-predicts the $K$-band flux.  However, we would expect differences in detail since the true geometry is unlikely to be the spherical shell assumed by the {\tt DUSTY} models.  However, two or three dimensional dust models are beyond our present
scope.

Such a two-component model does not explain the remaining near-IR emission.  We can fill in this emission by adding a $L=0.038L_\odot$, $T_{\ast}=3000$~K, $M=0.25M_\odot$ M-star, taken from a $10^7$~year PARSEC isochrone.  The combination of these three components provides a good semi-quantitative model of the SED at minimum.  The mismatch at $K$ band is not worrisome  because the spherical geometry of the DUSTY models  cannot be correct even if it provides a reasonable model of the quiescent star, and producing the occultation requires some change in geometry that presumably also produces some changes in the observed dust emission.  Dust emission probably cannot be used to explain the flux excess at the shorter near-IR bands because it requires material above the evaporation temperature of even graphitic dusts.

Introducing the M star is plausible for two additional reasons -- first, there is the evidence for periodicity (see Section \ref{subsec:orb}) and second, there has to be something driving the geometry change producing the occultation event, and
the simplest way of doing so is to make the system a binary.  We can also place approximate limits on stellar companions using the SED and the IR spectrum. A black body with a temperature and luminosity below the dashed black line in Figure \ref{fig:LT_iso} would not alter the optical SED of the dim phase by more than 20\%. And the shaded area shows where a black body can dilute the near-IR Br$\gamma$ line by 80\% while diluting the Pa$\beta$ line by less than 20\%. The M dwarf added to explain the occulted SED lies in the middle of this region.

\subsection{Orbital parameters}\label{subsec:orb}

Figure \ref{fig:all_lcs} shows all the available historical photometry of ASASSN-24fw.  After \cite{2024ATel16833....1J} discovered this transient, \cite{2024ATel16919....1N} suggested a period of 43.8 years (marked with arrows in Figure \ref{fig:all_lcs}) with an estimated duration of 8 months based on the DASCH light curve. They argue for possible events in 1937, when DASCH has 3 measurements and 14 upper limits below the standard deviation of the DASCH light curve, and in 1981 where there are 9 non-detections below this limit. To determine whether the event is periodic, we need to estimate the duration of the event. Figure~\ref{fig:lc_in} shows a cubic spline fit to the ASASSN/LCOGT/POISE/ROS2 $g$-band light curve (selecting only the ASAS-SN and ROS2 data brighter than 16 mag and with photometric errors below 1\%).  The source steadily fades until reaching a minimum on January 26, 2025.  It then brightened again and had almost completely recovered by the time it became unobservable at the beginning of June 2025.  If we reverse the spline fit around the time the star reached its minimum,  we see that the event is very nearly time symmetric and a slight shift of the central time by 4 days further improves the match of the reversed spline to the data.  If we define the event duration by the period when the source had faded by 0.2~(3.5) mag with respect to the quiescent magnitude, then the total duration is 261 (219) days. This is remarkably close to the 8-month duration suggested by \citet{2024ATel16919....1N}.

\begin{figure}
\centering
\includegraphics[width=\linewidth]{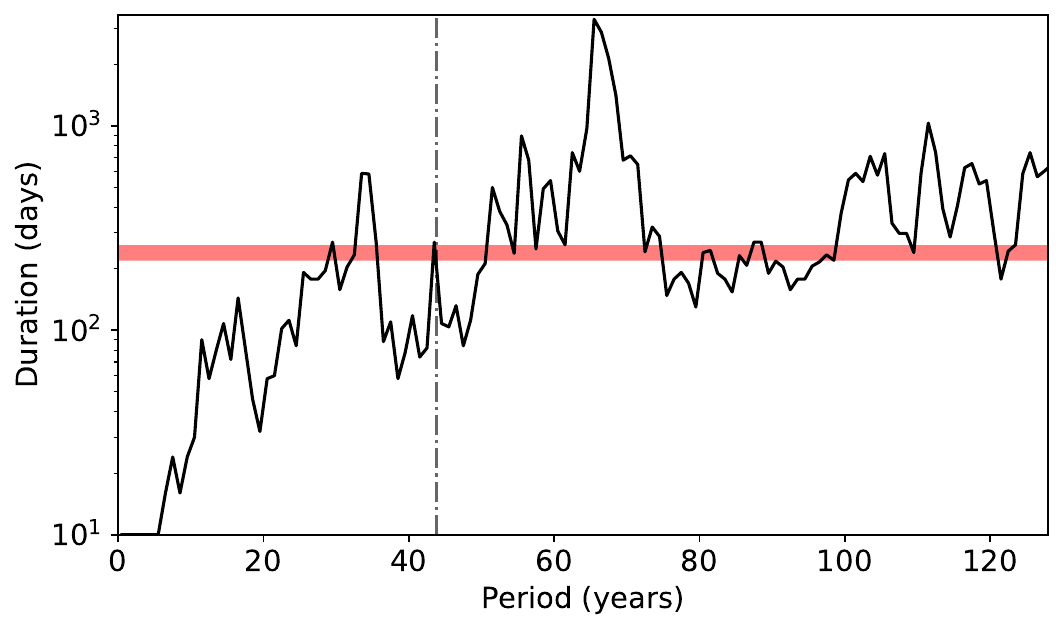}
\caption{Maximum allowed duration of missed events due to the gaps in the photometric coverage as a function of period. The red horizontal band mark the shallow (0.2 mag) and deep (3.5 mag) event durations and the dashed-dotted gray vertical line corresponds to the period proposed by \citet{2024ATel16919....1N}.\label{fig:period}}
\end{figure}

We tested possible periods by determining the longest duration event that would be permitted at each period given the gaps in the available light curves.  We dropped the DASCH epochs with lower limits or measurements below the standard deviation of the rest of the data so that events are allowed at these epochs. Figure \ref{fig:period} shows this maximum allowed duration as a function of period, where we have binned the periods and show only the largest allowed duration in each bin to suppress the noise from short duration solutions. The horizontal band shows the range of the transit duration estimates given above.  Periods shorter than 29 years are completely ruled out. The 43.8 year period suggested by \citet{2024ATel16919....1N} is marginally allowed, with a duration similar to the current event. 

\begin{figure}
\centering
\includegraphics[width=\linewidth]{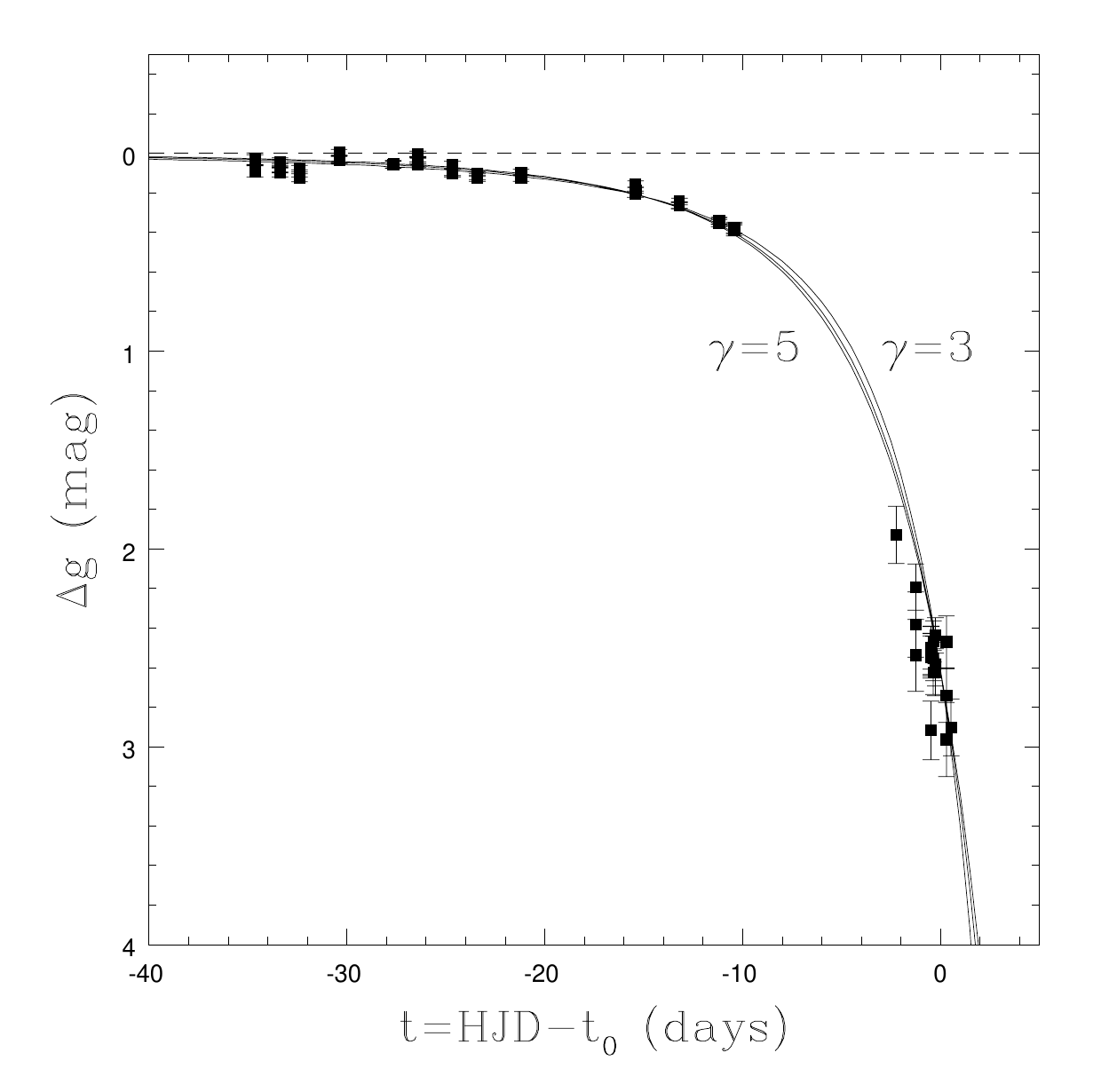}
\caption{Models of the ASAS-SN ingress light curve.  The points with errors are the ASAS-SN data, and the dashed line corresponds to the pre-event flux baseline.  The solid lines are the best models for optical depth exponents of $\gamma=3$, $4$ and $5$. \label{fig:lcfit}}
\end{figure}

\begin{figure}
\centering
\includegraphics[width=\linewidth]{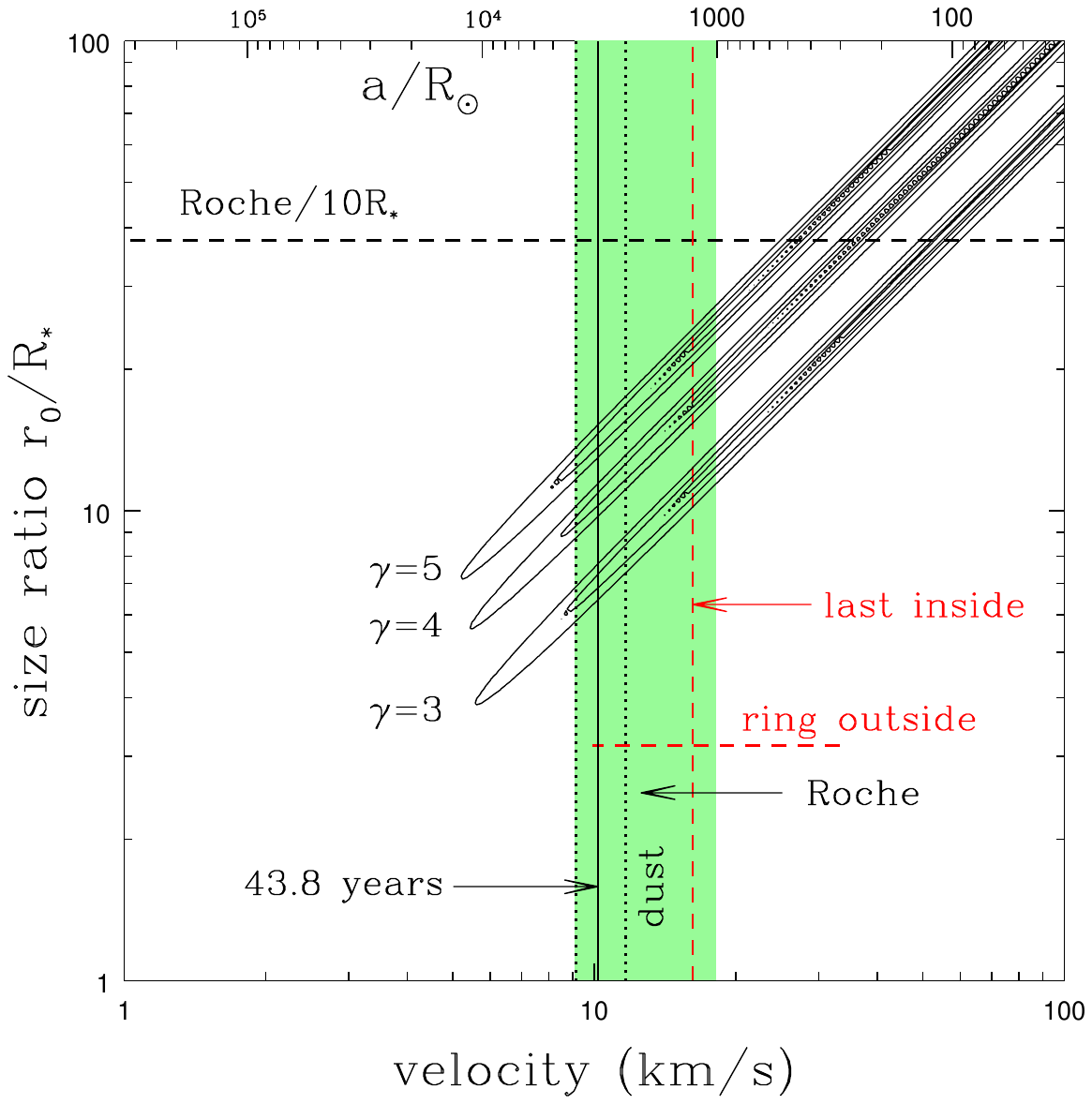}
\caption{Ingress constraints on the velocity $v$ and the optical depth scale length to stellar radius ratio $r_0/R_*$ for density exponents of $\gamma=3$ (bottom), $4$ and $5$ (top).  Contours are drawn at $\Delta\chi^2=1$, $3$, $10$ and $30$ after rescaling the minimum $\chi^2$ values to the number of degrees of freedom.  The axis at top converts the velocity into a semimajor axis assuming a circular orbit and a total mass of $1.70M_\odot$.  The heavy vertical black line is the orbit with the period proposed by \citet{2024ATel16919....1N}.  The green shaded region is the span of the  circumstellar dust in the model of the pre-event SED. The vertical dashed black lines are the limits of the M dwarf's binary Roche lobe ($a\pm R_R$) for this orbit. Any disk or other bound material orbiting the M dwarf must lie inside its Roche lobe. The horizontal dashed black line shows where the scale length of the density distribution is one tenth of the binary Roche radius ($r_0/R_*=R_R/10R_*$).  The vertical dashed red line labeled ``last inside'' marks the outermost orbit where the M dwarf could make a disk inside its orbit precess $180^\circ$ every 43.8~years.  The horizontal red dashed line labeled ``ring outside'' corresponds to the range of binary orbital semimajor axes that could make disk radii in the green shaded region precess $180^\circ$ every 43.8~years.
\label{fig:orbit}}
\end{figure}

To further constrain the orbit, we use the ingress to estimate a velocity. We fit a simple model where the optical depth is $\tau = \tau_0 (r/r_0)^{-\gamma}$, similar to the dust/debris disk models of \citet{2021AJ....161..238W} and \citet{2023MNRAS.519.5607R}. This model simply assumes that the optical depth near the edge of the occulter is a power law with negative exponent $\gamma$ in distance $r$ normalized by the optical depth $\tau_0$ at location $r_0$. We assumed we could neglect its curvature, and modeled the star as a square with sides of $2R_\ast=4.1 R_\odot$ so that we would only need a one dimensional integral. The change in magnitude from the baseline is then
\begin{equation}
    \Delta m(t) = -2.5\log\left[ \frac{r_0}{2R_\ast} \int_{u_-}^{u_+} du\exp\left(-\tau_0 u^{-\gamma} \right)\right]
\end{equation}
where
\begin{equation}
     u_\pm = 1 - \frac{v t}{r_0} \pm \frac{R_*}{r_0},
\end{equation}
$v$ is the relative velocity and the center of the star is at $r=r_0$ when $t=0$.
If $u$ is negative or the argument of the exponential too large, we set the flux contribution to zero.  The model has four parameters: $\tau_0$, $r_0$, $v$ and $\gamma$.  We fit the ASAS-SN $g$ band light curve minus the median pre-event magnitude, $\Delta g = g -13.11$, over the HJD time period 2460545 to 2460581 and dropped epochs which are only upper limits.  To avoid having three non-linear parameters, we averaged the 7 points taken over the last day of this time period 
to find $\Delta g_0 = 2.61 \pm 0.05$~mag at a mean HJD of $t_0=2460580.29$.  So we fixed $\tau_0 = 2.40$  and the time in the integral to $t=HJD-t_0$ so that at $t=0$ we will find a $\Delta m$ closely corresponding to $\Delta g_0$. We then computed the $\chi^2$ goodness of fit in the space of $r_0$ and $v$ for fixed $\gamma$.  Once $r_0/R_*$ is large, the model will be degenerate with $v \propto r_0$ because the size of the star is no longer important.

The goodness of fit for $\gamma=0.5$, $1$, $2$, $3$, $4$, $5$, and $6$ are $\chi^2=17479.2$, $980.5$, $306.1$, $ 206.7$ $196.2$, $203.2$, and $217.3$, respectively, for $61$ degrees of freedom.  This fit used the ``raw'' ASAS-SN errors, which are underestimates by a factor of $\sim 2$, so the best fits are actually quite good.  Shallow optical depth profiles (small $\gamma$) are strongly disfavored, and sharply rising profiles (large $\gamma$) are moderately disfavored. Figure~\ref{fig:lcfit} shows the three best models ($\gamma=3$, $4$ and $5$).  The fits trace both the first slow drop below the baseline near $t=-30$~days and the rapid drop as $t \rightarrow 0$.  While we could follow the decay further, it would require a model with more parameters to produce the flattish bottom of the event (see Figure~\ref{fig:lc_in}). Similar fits to the egress data were poor, possibly due to calibration problems produced by observing at extreme airmasses and the lack of a clear post-event baseline.

Figure~\ref{fig:orbit} shows the constraints on the velocity $v$ and scale length to stellar radius ratio $r_0/R_*$ for the same three values of $\gamma$. The best solutions basically lie along the degeneracy direction. While the formal $\Delta \chi^2$ values prefer the high $v$ and large $r/R_0$ solutions, this result is almost certainly very sensitive to the details of the calculation and the assumed profile.  Steeper profiles (larger $\gamma$) require larger scale lengths, as we would expect if the fits must produce the same light curve derivatives. Increasing $\gamma$ will steepen the
derivatives, which can be compensated by increasing $r_0$. We can also convert the velocity into a semi-major axis $a$ assuming a circular orbit and a total mass of $1.70 M_\odot$ (small changes to the total mass matter little given the logarithmic scales).  We also show the orbit with the $43.8$~year period proposed by \citet{2024ATel16919....1N}, and shade in green the region between the inner and outer dust radii from the pre-event SED model.

\subsection{Nature and origin of the occulter}\label{subsec:occ}

\begin{figure*}
\centering
\includegraphics[width=\linewidth]{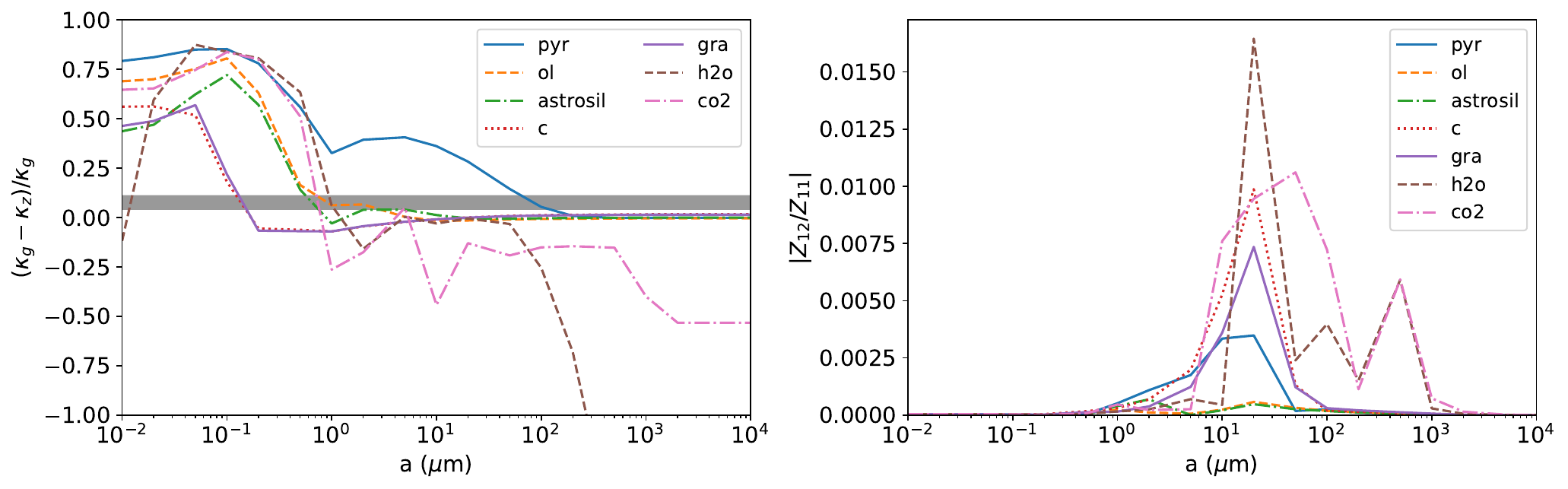}
\caption{The left panel shows the relative color change that different materials produce between the $g$ and $z$ bands as a function of particle size. The shaded band is the measured color change of the event. The right panel shows the ratios between Stokes Q and I in $V$ band for one forward scattering.\label{fig:op-pol}}
\end{figure*}

The optical depth of the circumstellar dust detected prior to the event ($\tau_V=0.25$, see Section \ref{subsec:star}) is too low to cause the 4.1 mag dimming (it requires at least $\tau_V=3.8$ if the star is fully covered, see Table \ref{tab:depth}), so there must be another dust source producing the polarized achromatic occultation. A purely stellar eclipse is also ruled out (see Section \ref{subsec:ecl}).

We can constrain the occulter properties by (1) the (near) achromaticity of the event in the optical and (2) the need to produce significant polarization. The slight correlation between $V$ magnitude and polarization seems to indicate that the polarization will decrease as the event ends and possibly be negligible in the quiescent phase of the star. To compare with observations, we compute the effective opacity, $\kappa=\sqrt{\kappa^{\text{abs}}(\kappa^{\text{abs}}+\kappa^{\text{scat}})}$, at the central wavelengths of the $g$ and $z$ bands (the bluest and reddest filters with REFCAT and LCOGT measurements) and compute the relative change, $(\kappa_g-\kappa_z)/\kappa_g$. This corresponds to the relative color change $\Delta (g-z)/\Delta g=0.08\pm0.04$. There is no simple expression for the polarization fraction, but the ratio between the scattering matrix elements $Z_{12}(\theta=0)$ and $Z_{11}(\theta=0)$ gives an idea of the ability of the material to polarize incoming natural light ($I_0$) with one forward ($\theta=0$) scattering. The transmitted intensity will be $I\propto Z_{11}I_0$ and the linearly polarized light will be $Q\propto Z_{12}I_0$ (for randomly oriented or axisymmetric grains, one scattering does not produce a Stokes U component). Thus, the ratio $|Z_{12}/Z_{11}|$ at the center of the $V$ band gives an estimate of the ability of a grain to produce polarized light.

We use \texttt{optool} \citep{2005A&A...432..909M,2016A&A...586A.103W,2021ascl.soft04010D} to compute the opacities and scattering matrices of pyroxene \citep[Mg$_{0.7}$Fe$_{0.3}$SiO$_3$ labelled as pyr,][]{1995A&A...300..503D}, olivine \citep[MgFeSiO$_4$ labelled as ol,][]{1995A&A...300..503D}, silicate grains \citep[astrosil,][]{2003ApJ...598.1017D}, carbon \citep[c,][]{1996MNRAS.282.1321Z}, graphite \citep[gra,][]{2003ApJ...598.1026D}, water ice \citep[h2o, ][]{2008JGRD..11314220W} and carbon dioxide ice \citep[co2, ][]{1986ApOpt..25.2650W}. We consider a range of particle radii from 0.01 to 10000 $\mu$m.

Figure \ref{fig:op-pol} shows the chromaticity, $(\kappa_g-\kappa_z)/\kappa_g$, and polarizability, $|Z_{12}/Z_{11}|$, as a function of grain radius, $a$, for the seven compositions. From this Figure we see that pyroxene dust can only produce achromatic dimming if the dust particles are larger than $\sim$100~$\mu$m, but such large pyroxene particles are not good sources of polarized emission. Silicates (olivine and astrosil) can produce the observed achromaticity with particles larger than $\sim$1~$\mu$m, but they also do not significantly polarize light. Carbonaceous dusts (carbon and graphite) are able to produce the required achromaticity with particles as small as $\sim$0.1~$\mu$m and both are good at producing polarization with particle sizes of $\sim$20~$\mu$m. Water ice can avoid significant color changes and produce significant polarization for a$\sim$10-50~$\mu$m, but larger and smaller gains lead to large color changes. While carbon dioxide ice is good at polarizing, it produces strong chromaticity. Thus, the occulter is likely composed of large ($\sim$20~$\mu$m) carbonaceous or water ice particles.

If we assume that the minimum velocity of the occulter is in the order of 10~km/s (see Figure~\ref{fig:orbit}) and use the 219-day duration of the deep eclipse, we can estimate a lower limit of the occulter mass. If the composition of the occulter is $20$~$\mu$m carbon dust particles, the mass of the occulter should be $M=\pi D^2 \Sigma/4=\pi D^2 \tau/4\kappa\gtrsim1.8\times10^{-3}M_{\oplus}$ for the diameter estimate of the occulter, $D\gtrsim$ 272~R$_\odot$ (or, equivalently, 1.27~AU), and a surface density, $\Sigma=\tau/\kappa$, given by the required optical depth and an effective opacity of $\kappa=403$ cm$^2$/g for the $g$ band. If the occulter is primarily made of $a=20$~$\mu$m water ice ($\kappa=13$ cm$^2$/g), the mass of the occulter would be at least $M\gtrsim0.055M_{\oplus}$.

F stars do not have dusty winds, but even if we arbitrarily give the star a large mass loss rate, such large dust grains cannot have been formed in a stellar wind from the observed star.  Dust in a wind forms when the temperature of a forming grain is less than the condensation temperature $T_c = 1500 T_{c0}$~K. This leads to a formation radius of
\begin{equation}
       R_f = \left( \frac{ L_*}{16 \pi \sigma T_c^4 } \right)^{1/2}
           \simeq 23 L_{*10}^{1/2} T_{c0}^{-2}  R_\odot
\end{equation}
for a stellar luminosity of $L_* = 10 L_{*10} L_\odot$. Once grains begin to condense, they grow
by collisions to an asymptotic size of
\begin{equation} 
   a_f = \frac{ v_c f \dot{M}}{16 \pi \rho_b v_w^2 R_f}
     \simeq 0.1 \frac{ f_0 \dot{M}_4 v_{c1} T_{c0}^2 }{  L_{*10}^{1/2} \rho_0 v_{w4}^2}~\mu\hbox{m}
\end{equation}
where $v_c = v_{c1}$~km/s is the effective collision velocity, $f=0.005 f_0$ is the mass fraction of the condensible
species, $\dot{M} = 10^{-4} \dot{M}_4$~$M_\odot$/year is the mass loss rate, the wind velocity ($v_w = 400 v_{w4}$~km/s) is scaled to the estimated escape velocity of the star, and $\rho_b = 3 \rho_0$~g~cm$^{-3}$ is the bulk grain density (see, e.g., \citealt{Kochanek2014}).  Even when scaled to the extreme mass loss rate of an AGB star ($10^{-4}$~$M_\odot$/year), the resulting grains are far too small, in large part because the characteristic velocity of a wind is the escape velocity, which is large. 

A fairly steady mass loss rate is also required.  If we imagine a brief, impulsive mass loss event that has the potential to reach even more extreme densities, the dust forms in an expanding shell, leading to an optical depth that must drop at least as fast as $t^{-2}$ (see, e.g., \citealt{Adams2014}), making it impossible to have the slowly varying and symmetric observed obscuration. Taken together, these problems seem to require some other origin for the dust than the F star.

If the star is young, its protoplanetary disk may be dominated by large dust grains due to grain growth during the protostellar phase \citep{2006ApJ...644L..71S,2014prpl.conf..339T,2022MNRAS.515.4780T}. \citet{2003A&A...412L..43P} found evidence in T Tauri disks for silicate grains up to 2~$\mu$m in size and \citet{2005ApJ...622..404K} detected an absence of warm ($T=300$~K) silicate grains smaller than 10~$\mu$m in T Tauri stars. Both works were focused on silicate dusts, and the behavior of carbonaceous and ice materials under these conditions is unexplored.

Instead of a protoplanetary disk, the occulter could be a dwarf star or an exoplanet with extended rings. For example, Saturn's rings also have a power law distribution of particle sizes, with $dn/d\ln a \sim a^{-2}$ (\citealt{Cuzzi2009}) with a dearth of particles smaller than
$ a \simeq 100$-$1000~\mu$m (\citealt{Ohtsuki2020}).  Although the composition would have to be different from Saturn's water ice dominated rings since such large water particles would not be sufficiently achromatic.

The long and deep occultation of ASASSN-21qj was explained by a collision event  \citep{2023ApJ...954..140M,2023Natur.622..251K}. In that case, an infrared brightening was detected 2.5 years earlier than the dimming and the eclipse was irregular and chromatic. In contrast, ASASSN-24fw showed no changes in its mid-IR brightness during the full period of NEOWISE observations (see Figure \ref{fig:all_lcs}) and the in-transit light curve is mostly flat (see Figure \ref{fig:lc_in}), ruling out a recent collision event. Unfortunately, the NEOWISE observations ended shortly before the start of the event.

Alternatively, the occulter might be produced by an old collision of bodies (e.g. planets, asteroids or comets). Such collisions typically produce fragments with a size distribution of $dn/d\ln a \propto a^{-2.5}$ \citep[see, e.g.,][]{Dohnanyi1970,1986MmSAI..57...47F} which is dominated by small grains that will produce large color changes. However, if the debris forms a steady-state disk, radiation pressure from the star will remove the smallest grains to leave a debris disk dominated by grains with sizes on the order of tens of $\mu$m, although the smallest grains are a few $\mu$m \citep{2006A&A...455..509K}. This also avoids the dynamical evolution of the optical depth expected during a recent collision and it is probably needed to have a light curve with the observed degree of symmetry.

\subsection{System geometry}\label{subsec:geom}

\begin{figure*}
\centering
\includegraphics[width=\linewidth]{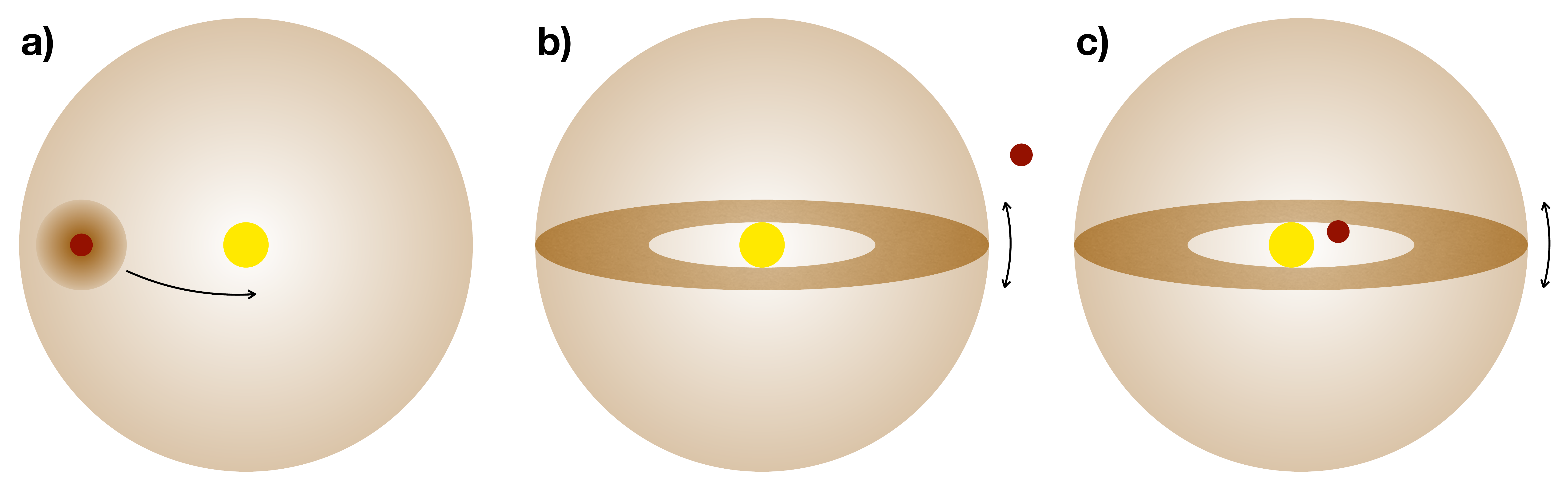}
\caption{Schematic diagrams of the possible configurations discussed for ASASSN-24fw. They are not drawn to scale. The yellow circle is the F star, the dark red circle is the proposed M dwarf companion and the light brown sphere is the optically thin circumstellar dust. In the scenario a) the M dwarf with a dusty disk (dark brown) is in a 43.8 year period around the F star and the disk occults the F star. In b) an optically thick disk (dark brown disk) precesses around the F star due to the M star orbiting outside the circumstellar disk. In the last scenario, c) the optically thick disk lies outside the binary orbit and the precession is driven by the interior binary.\label{fig:conf}}
\end{figure*}

Based on the observations presented, we propose that the system has three components: a 7$L_{\odot}$ F-type star (that can be either young or slightly evolved), a dust structure formed by fairly large ($\sim20$~$\mu$m) particles and a low-mass companion.  We discuss three possible system geometries for producing a 43.8~year periodicity without invoking additional components and schematically show the configurations in Figure \ref{fig:conf}. First, we explore making it the orbital period of the binary.  Then we discuss making it a disk precession period with the disk either around the F star interior to the binary orbit, or as a circumbinary disk.

In Figure~\ref{fig:orbit}, we see that if we place the M dwarf in a 43.8~year orbit around the F star, it would lie where the SED model puts the circumstellar dust, see Figure \ref{fig:conf}a). Using the \cite{1983ApJ...268..368E} approximation and the orbit inferred in Section~\ref{subsec:orb}, the binary Roche radius of the M dwarf is $R_R=770 R_\odot$.  The time to cross the Roche radius for the $43.8$~year period is $R_R/v \simeq 620$~days, so a dust distribution inside the binary Roche lobe and extending to $R_d \sim R_R/4 \sim 190 R_\odot$ could produce an event lasting $\sim 300$~days. The  scale length ratio in the optical depth model of $r_0/R_* \sim 10$ means that the gradients are also a small fraction of the binary Roche radius.  We can also ask where the dust  heating by the F star equals that by the M star. For large grains, the Planck absorption factors will be the same for both stars, so this is just the radius
$\sim 220 R_\odot$ from the M dwarf where the fluxes are equal.  This means that the heating of the dust producing the occultation is dominated by the M dwarf.

The dust associated with the M dwarf cannot produce the observed mid-IR excess.  Even if all of the F star flux impinging on the Roche radius was absorbed, it would represent only 5\% of the F star flux.  If all of the M dwarf's emission was also absorbed, this would only produce a total dust luminosity of $0.4 L_\odot$.  However, the high optical depth region around the M star appears to be significantly smaller, $R_d\sim R_R/4$, which would only capture $0.2\%$ of the F star's flux, leading to a total dust luminosity of only $0.05L_\odot$ if
all the M dwarf flux was also absorbed.  This is well below the dust emission seen in Figure~\ref{fig:SED}.  Moreover, the pre-occultation SED model also needs some other source of dust absorption to fit the optical SED.

There are several problems with associating the dust with the M dwarf.  Figure~\ref{fig:orbit} also shows where the M dwarf's Roche radius would lie relative to the orbit, so the gravitational sphere of influence of the dwarf encompasses much of the radial range of the diffuse dust and
the M dwarf will develop a dust overdensity simply by gravitationally focusing the diffuse dust (a version of Problem 4.4
in \citealt{Binney2008}).  This will (very roughly) produce an optical depth excess of order 
\begin{equation}
   \tau \sim \frac {\tau_d}{ \Delta R }\left(\frac{G M_d R_d}{\sigma_0^2}\right)^{1/2}
         \sim 0.2 \left(\frac{M_d}{0.25M_\odot}\frac{R_d}{100R_\odot}\right)^{1/2}
                \frac{\hbox{km/s}}{\sigma_0}
\end{equation}
where $M_d$ is the mass of the object that produces the density enhancement (i.e., the M dwarf), $R_d$ is the size of the density enhanced region, $\sigma_0$ is the dispersion of the dust particle velocities compared to the orbital velocity, and $\Delta R$ is the width of the region over which the diffuse dust producing the optical depth $\tau_d$ in the SED is spread. 
This scale is multiplied by an complex function of $R/R_d$ that slowly rises as the distance $R$ of the sight line to the dwarf becomes smaller. If it was possible to gain a factor of 20, then this mechanism would produce a cloud of dust around the dwarf that could do the  occultation, although the effects of the gravity from the F star would need to be included in a fully quantitative model.   However, the cloud of dust would be (quasi-)spherical, so the M dwarf 
would lie at the center and have a higher optical depth than the line of sight to the F star, which would make it difficult to use the M dwarf to fill in the near-IR fluxes.

This latter problem could be avoided if the dust could be captured into a disk, which requires collisions/dissipation.  For a cloud of just dust, the collisional rates are negligible and a disk cannot be formed.  If it were a cloud of dust plus gas, then the gas could provide the dissipation to form a disk.  This would then need to be inclined relative to the orbital plane both in order to produce a long duration event and to avoid also obscuring the dwarf.  However, it is the orbital plane that sets the orientation of the angular momentum, so it is not obvious how capture of dust into a disk also produces a disk with a significant inclination.  Overall, trying to associate the disk with the M dwarf with out a mechanism for producing the disk does not seem promising. 

The other possibility is that there is an extended disk of dust in the system with a higher density than the lower density material required by the SED model. If this disk is inclined relative to the orbital plane of the binary, then precession can be used to drive the periodicity.
There are two possibilities: (1) the dust disk is around the F star and inside the orbit of the dwarf, Figure \ref{fig:conf}b); and (2) that the dust disk is circumbinary, Figure \ref{fig:conf}c).  We consider both cases.  In both cases we assume that the observed recurrence time $t_p=43.8$~years, corresponds to precession by $180^\circ$.

For a disk around the F star and inside the M dwarf orbit, we use the precession time from \citet{1996MNRAS.282..597L},
\begin{equation}
    \frac{\pi}{t_p} = -\frac{3G M_s}{4 a^3}\frac{P_d}{2\pi}\cos \delta,
\end{equation}
where $M_s= 0.25 M_\odot$ is the mass of the secondary, $a$ is the semimajor axis of the binary orbit, $P_d$ is the period of the disk orbit treated simply as a ring, and $\delta$ is the inclination of the binary orbit with respect to the initial disk mid-plane.  We can solve for the period of the binary
\begin{equation}
   P_b = \left( \frac{3}{2}\frac{M_s}{M_T} P_d t_p \cos\delta \right)^{1/2}
               = 3.1 P_d^{1/2} \cos^{1/2} \delta~\hbox{years}
\end{equation}
assuming a circular binary orbit where $M_T=1.70M_\odot$ is the total mass of the stars and orbital period of the disk $P_d$ is also in years in the numerical expression.  At the period
\begin{equation}
   P_0 = \frac{3}{2}\frac{M_s}{M_T} t_p \cos\delta = 9.7\cos\delta~\hbox{years}
\end{equation}
the condition $P_b=P_d=P_0$ is fulfilled and having the ring inside the orbit requires $P_b<P_0$.  This largest orbit lies just inside the region where the SED model places the dust (see vertical red line in Figure~\ref{fig:orbit}).  Hence, it seems difficult or impossible to 
place the disk fully inside the binary orbit.

For a circumbinary disk, we use the approximation from \citet{2004ApJ...607..913C} that gives
\begin{equation}
  \frac{\pi}{t_p} = -\frac{2\pi}{P_d}\left(\frac{a_b}{a_d}\right)^2,
\end{equation}
where $a_b$ and $a_d$ are the semimajor axes of the binary and the disk, respectively.  We can again solve for the binary period in terms of the disk period
\begin{equation}
    P_b = P_d^{7/4} \left( 2 t_p\right)^{-3/4} = 0.035 P_d^{7/4}~\hbox{years}
\end{equation}
where in the numerical result, $P_d$ is again in years.  Alternatively, we
can solve for the semimajor axis of the binary
\begin{equation}
   a_b = a_d \left( \frac{\pi^2 a_d^3}{G M_T t_p^2}\right)^{1/4}
     = 296 \left( \frac{a_d}{1000 R_\odot}\right)^{7/4} R_\odot.
\end{equation} 
The length of the horizontal red line in Figure~\ref{fig:orbit} shows the range of the binary semimajor axes 300 $R_\odot$ to 3500 $R_\odot$ corresponding to
having $a_d$ span the radial range of the dust in the SED model.  It is now
relatively easy to have much of the dust lie outside the binary orbit
and have a $2 \times 48.3$ year precession period. In particular, if the binary semimajor axis is between $\sim$ 300 to 1000 $R_\odot$, it can be completely inside the circumstellar dust and make the dusty disk precess with the required period. Although a circumbinary disk need not produce symmetric eclipses, there are other examples in the literature where a circumbinary disk produces symmetric eclipses such as KH 15D \citep{2006ApJ...644..510W,2021MNRAS.503.1599P} or Bernhard-1 and Bernhard-2 \citep{2022ApJ...933L..21Z}. Given the shape of the eclipse and the precession period constraint, this seems the most promising scenario at this point.

\section{Population analysis of long deep eclipses}\label{sec:pop}

\begin{sidewaystable*}
\vspace{-10cm}
\caption{Summary of long and deep eclipses. We include the name, {\it Gaia} ID, distance, $E(B-V)$ extinction, duration, period, depth and known references for each system.\label{tab:ecl}}
\begin{adjustbox}{width=1\textwidth,center}
\begin{tabular}{l l r@{}l c r@{}l c c c}
\hline\hline
Name & {\it Gaia} DR3 ID & \multicolumn{2}{c}{Distance (pc)}& $E(B-V)$ (mag) & \multicolumn{2}{c}{Duration (days)} & Period (years) & Depth (mag) & References \\ 
\hline
OGLE-LMC-ECL-17782 & 4658430243457185664 & & & & 2&.67 & 0.037 & 0.4 in $I$ & (1) \\ 
ASASSN-V J213939.29$-$702817.3 & 6396259296582909440 & 1053&$_{- 15 }^{+ 11 }$ & 0.026 & 3&.1 & Unknown & $\sim$1.3 in $g$ & (2) \\ 
CHS 7797 & No {\it Gaia} ID & && & $\sim$12& & 0.0487 & $\sim$1.7 in $R$, $I$, and $z'$ & (3) \\ 
OGLE-LMC-ECL-11893 & 4658294458116546304 & 7500&$_{- 1500 }^{+ 2600 }$ & 0.920 & $\sim$15& & 1.281 & Irregular, up to 1.5 in $I$ & (4) \\ 
PDS 110 & 3220462655745525632 & 343&$\pm3$ & 0.115 & 25& & 2.21 & 0.33 in $V$ & (5) \\ 
EPIC 204278916 & 6243130106031671168 & 139.6&$\pm0.4$ & 0.000 & $\sim$25& & Unknown & Irregular, up to 1.1 in Kepler band & (6) \\ 
Bernhard-2 & 3048281986701404160 & 2900&$_{- 500 }^{+ 1300 }$ & 0.301 & 26& & 0.172 & 2.45, 2.12, 1.85, 1.54, 1.57 in $g/r/i/z/y$ & (7) \\ 
VVV-WIT-07 & 5974962995291907584 & 5200&$_{- 1800 }^{+ 2200 }$ & 2.118 & $\sim$50& & 0.46 or 0.88 (tentative) & 1.75 in NIR & (8) \\ 
V1400 Cen & 6117085769513415168 & 137.9&$\pm0.2$ & 0.024 & $\sim$54& & Unknown & Irregular, up to 3 in $V$ & (9) \\ 
ASASSN-21co & 6647970630972147840 & 2660&$_{- 110 }^{+ 130 }$ & 0.109 & 66& & 11.9 & 0.6 in $V$ & (10) \\ 
ELHC 10 & 4658166433748357760 & 18000&$_{- 4000 }^{+ 5000 }$ & 1.139 & $\sim$66& & 0.602 & $\sim$0.6 in $I$ & (11) \\ 
VSSG 26 & No {\it Gaia} ID & && & $\sim$70& & 0.3583 & 0.4 in $K_s$ & (12) \\ 
YLW 16A & No {\it Gaia} ID & && & $\sim$80& & 0.2536 & 0.95 in $K_s$ & (13) \\ 
OGLE-BLG182.1.162852 & No {\it Gaia} ID & && & 100& & 3.496 & Variable, between 2 to 3 in $I$ & (14) \\ 
Bernhard-1 & 2061078599753657728 & 2300&$_{- 400 }^{+ 500 }$ & 0.780 & 112& & 0.526 & 2.19, 2.12, 1.53, 1.47, 1.46 in $g/r/i/z/y$ & (7) \\ 
$\eta$ Geminorum & 3377072212924335488 & && & 150& & 8.2 & 0.4 in $V$ & (15) \\ 
V773 Tau & 163184366130809984 & 119&$\pm2$ & 0.000 & 150& & 26.5 & 1.3 in $V$ & (16) \\
WD J1237+5937 & 1578454838386128384 & 193&$\pm16$ & 0.000 & 164& & Unknown & 0.42 achromatic & (17) \\ 
VVV-WIT-08 & 4044152029396602880 & 8500&$\pm1600$ & 0.633 & $\sim$200& & Unknown & 3.8 achromatic & (18) \\ 
ASASSN-24fw & 3152916838954800512 & 1011&$_{- 23 }^{+ 15 }$ & 0.062 & 261& & 43.8 (suggested) & 4.12, 3.95, 3.81, 3.76, 3.39, 2.84, 1.9 in $g/r/i/z_s/J/H/K_s$ & This work, (19) \\
WD J1013$-$0427 & 3780094656734582528 & 304&$_{- 12 }^{+ 10 }$ & 0.035 & 489& & Unknown & 0.38, 0.27, 0.20, 0.20 in ZTF $g/$ATLAS $c/$ZTF $r/$ATLAS $o$ & (17) \\
ASASSN-21qj & 5539970601632026752 & 557&$\pm3$ & 0.364 & $\sim$500& & 63 (predicted) & Irregular, up to $\sim$3 in $g$ & (20) \\
ASASSN-V J060000.76$-$310027.83 & 2891196718939580672 & 155.8 &$_{-0.2}^{+0.3}$ & 0.000 & $\sim$580& & Unknown & 0.9 in $g$ & (21) \\
V409 Tau & 150393571269837184 & 129.5&$\pm0.4$ & 0.000 & $\lesssim$630& & Unknown & $\sim$1.4 in $V$ & (22) \\ 
$\varepsilon$ Aurigae & 205499655242974464 & 1060&$_{- 190 }^{+ 370 }$ & 0.018 & $\sim$668& & 27.1 & $\sim$0.75 achromatic & (23) \\ 
WD J1302+1650 & 3937407901354145536 & 400&$_{- 14 }^{+ 17 }$ & 0.000 & 691& & Unknown & 0.27 achromatic & (17) \\
Gaia21bcv & 3045209156636885760 & 1360&$\pm80$ & 0.981 & 866& & Unknown & Irregular, up to 3 in {\it Gaia} G and ZTF $r$ & (24) \\ 
M 2-29 & 4063244773875284224 & 6600&$_{- 900 }^{+ 1900 }$ & 0.521 & $\sim$1000& & Unknown & 1.3 in I, 1.7-1.8 in $V$ & (25) \\ 
TYC 2505-672-1 & 795188391420888192 & 1370&$_{- 40 }^{+ 30 }$ & 0.000 & 1260& & 69.1 & 4.5 in the optical & (26) \\ 
V718 Per & 216678115082741248 & 307&$\pm3$ & 1.874 & 1278& & 4.7 & 0.66 in $I$ & (27), (28) \\ 
J202402+383938 & 2061158279988516224 & 1830 &$\pm50$ & 0.653 & 2200& & Unknown & 0.59 in $g$ & (29) \\
Gaia17bpp & 1824311891830344704 & 8500&$_{- 1700 }^{+ 2300 }$ & 1.312 & 2374& & Unknown & 4.5 in {\it Gaia} $G$ & (30) \\  
ASASSN-21js & 5334823481651325440 & 2610&$_{- 90 }^{+ 100 }$ & 0.385 & Ongoing, 2746& \,(predicted) & 610000 (predicted) & 0.24 in $g$ & (31) \\ 
ASASSN-24fa & 5697179770616369664 & 4800&$\pm300$ & 0.187 & Ongoing, $>$1220& & Unknown & 0.3 in $g$ & (32) \\ 
AA Tau & 147818450613367424 & 133.3&$_{- 1.4 }^{+ 2.3 }$ & 0.000 & Ongoing, $>$5300& & Unknown & Irregular, up to $\sim$2 in $V$ & (33) \\ 
WeSb 1 & 423384961080344960 & 3200&$_{- 200 }^{+ 300 }$ & 0.239 & Variable,& \,few days & Aperiodic, 1 (tentative) & Irregular, up to 3 achromatic & (34), (35) \\
``Tabby’s Star" & 2081900940499099136 & 434.3&$_{- 2.0 }^{+ 1.7 }$ & 0.314 & \multicolumn{2}{c}{Variable, between 5 to 80} & Aperiodic & Irregular, up to $\sim$0.24 in Kepler band & (36) \\
ZTF J0139+5245 & 407197396840413696 & 178 &$_{- 5 }^{+ 6 }$ & 0.000 & \multicolumn{2}{c}{Variable, between 15 to 25} & 0.29 & Irregular, up to 0.6 in ZTF $r$ & (37) \\
ASASSN-25bv & 5332048241235589504 & 1030 &$\pm70$ & 0.961 & \multicolumn{2}{c}{Variable, between 27 to $>$110 (ongoing)} & Unknown & Irregular, up to 0.9 in $g$ & (38) \\
ZTF J185259.31+124955.2  & 4506139331756845568 & 4700&$_{- 700 }^{+ 900 }$ & 0.471 & Variable,& \,$\sim$40 & 0.79 & 0.4 in ZTF $g/r$ and ATLAS $c/o$ & (39) \\
ZTF J0347$-$1802 & 5107322396824711680 & 75.6 &$_{- 0.7}^{+ 0.6}$ & 0.000 & Variable,& \,up to 70 & Aperiodic & Irregular, up to $\sim$0.3 in ZTF $r$ & (40) \\
SBSS 1232+563 & 1571584539980588544 & 172&$\pm3$ & 0.000 & Variable,& \,up to 240 & Aperiodic & Irregular, up to 0.75 achromatic & (41)	\\
TIC 400799224 & 5238414793089292160 & 700&$_{- 100 }^{+ 180 }$ & 0.241 & Variable& & $\sim$0.0541 & Irregular & (42) \\ 
KH 15D & 3326686439745822336 & 730&$\pm30$ & 1.078 & Variable& & 0.131 & Variable, up to 5 in $I$ & (43), (44) \\ 
EE Cephei & 2197941958898810240 & 2200&$_{- 70 }^{+ 90 }$ & 0.289 & Variable& & 5.6 & Variable, between 0.5 to 2.0 in the optical & (45) \\
ZTF J0923+4236 & 817461778282929664 & 147.7 &$_{- 1.8 }^{+ 1.4 }$ & 0.000 & Variable& & Aperiodic & Irregular, up to $\sim$0.5 in ZTF $r$ & (40) \\
TYC 8830-410-1 & 64916065741212997121 & 158.0&$_{- 0.4 }^{+ 0.3 }$ & 0.008 & Variable& & Aperiodic & Irregular, up to $\sim$1 in $V$ & (46) \\ 
\hline
\end{tabular}
\end{adjustbox}
\newline

{\footnotesize References: (1) \citet{2011AcA....61..103G}, (2) \citet{2019ATel12836....1J}, (3) \citet{2012AA...544A.112R}, (4) \citet{2014ApJ...788...41D}, (5) \citet{2017MNRAS.471..740O}, (6) \citet{2016MNRAS.463.2265S}, (7) \citet{2022ApJ...933L..21Z}, (8) \citet{2019MNRAS.482.5000S}, (9) \citet{2012AJ....143...72M}, (10) \citet{2021RNAAS...5..147R}, (11) \citet{2016MNRAS.457.1675G}, (12) \citet{2008ApJ...684L..37P}, (13) \citet{2013AA...554A.110P}, (14) \citet{2015MNRAS.447L..31R}, (15) \citet{2022MNRAS.516.2514T}, (16) \citet{2022AA...666A..61K}, (17) \citet{2025PASP..137g4202B}, (18) \citet{2021MNRAS.505.1992S}, (19) \citet{2024ATel16919....1N}, (20) \citet{2023Natur.622..251K}, (21) \citet{2019ATel13346....1W}, (22) \citet{2015AJ....150...32R}, (23) \citet{2010AJ....139.1254S}, (24) \citet{2024AJ....167...85H}, (25) \citet{2008AA...490L...7H}, (26) \citet{2016AJ....151..123R}, (27) \citet{2003ApJ...596L.243C}, (28) \citet{2006ApJ...646L.151N}, (29) \citet{2025arXiv250719594J}, (30) \citet{2023ApJ...955...69T}, (31) \citet{2024AA...688L..11P}, (32) \citet{2024ATel16715....1J}, (33) \citet{2021AJ....161...61C}, (34) \citet{2025PASP..137b4201B}, (35) \citet{2025NatAs...9..380B}, (36) \citet{2016MNRAS.457.3988B}, (37) \citet{2020ApJ...897..171V}, (38) \citet{2025ATel17196....1J}, (39) \citet{2024AA...688A..58B}, (40) \citet{2021ApJ...912..125G}, (41) \citet{2025ApJ...980...56H}, (42) \citet{2021AJ....162..299P}, (43) \citet{2006ApJ...644..510W}, (44) \citet{2021MNRAS.503.1599P}, (45) \citet{2020AA...639A..23P}, (46) \citet{2021ApJ...923...90M}.}

\end{sidewaystable*}

\begin{figure*}
\centering
\includegraphics[height=0.4\textheight]{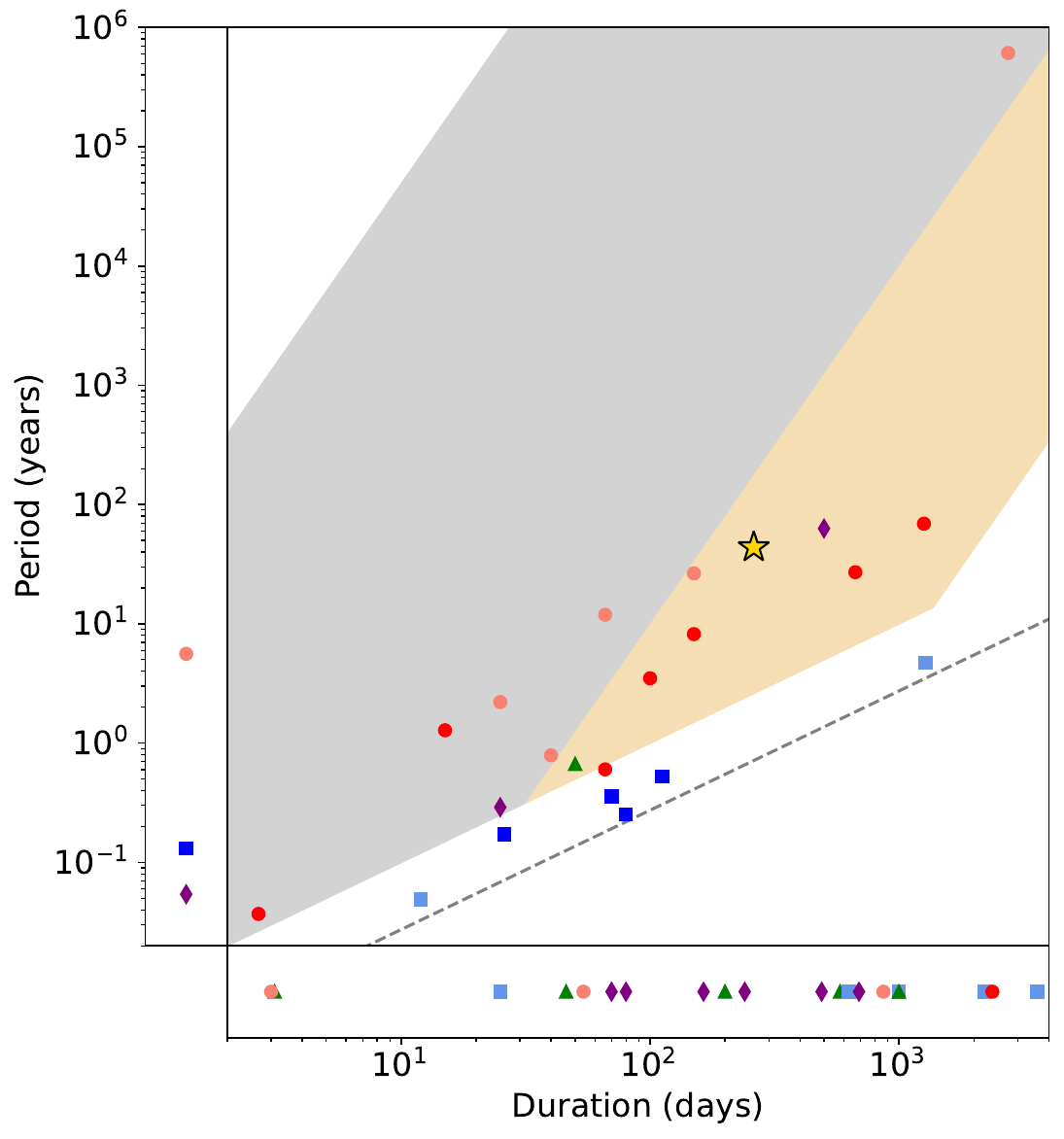}
\hfill
\includegraphics[height=0.4\textheight]{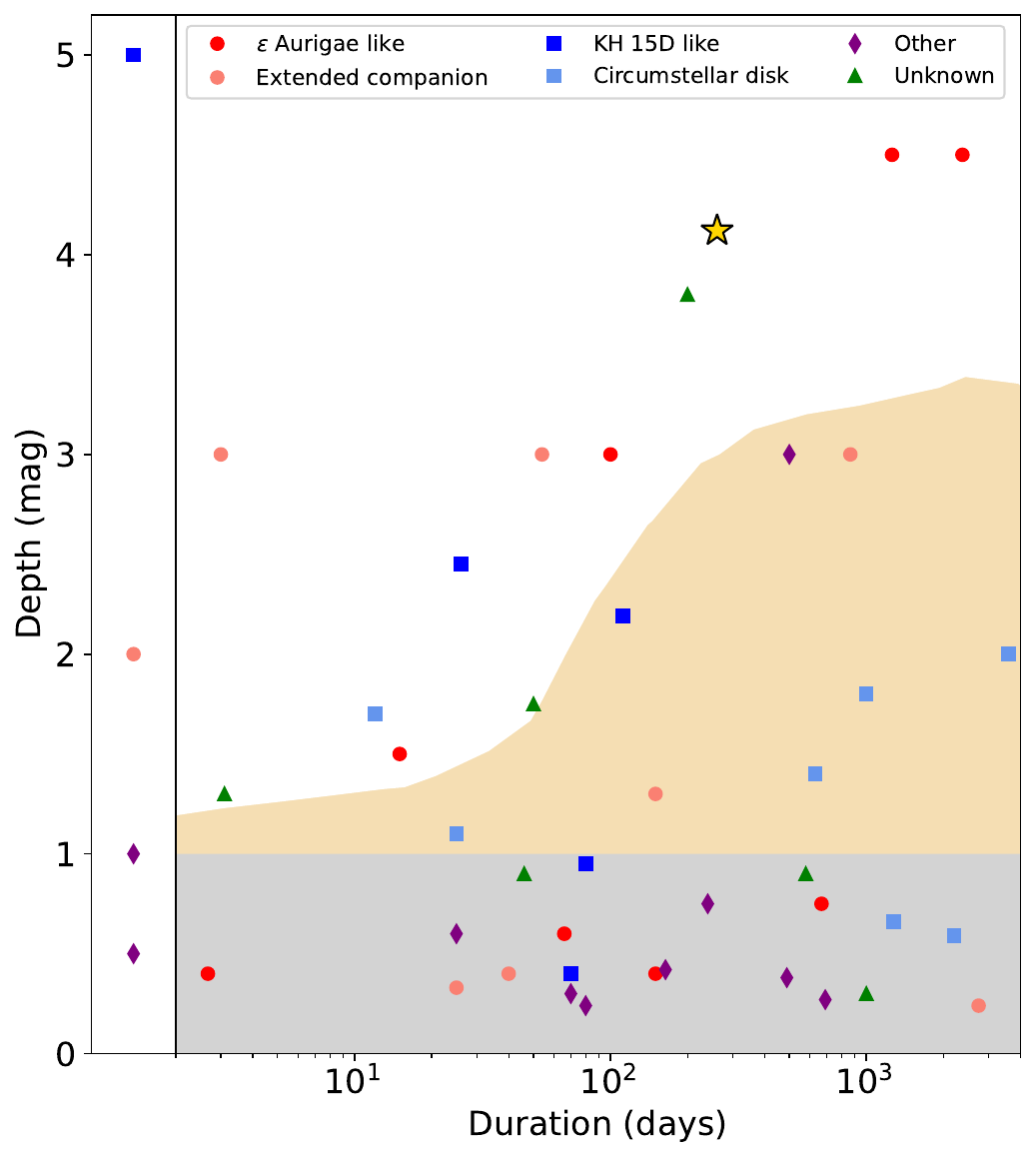}
\caption{Period (left) and depth (right) against duration for the long and deep eclipses. The systems on the side (bottom) subpanel have variable or unknown duration (period). The red circles are systems like $\varepsilon$ Aurigae. The light red circles are systems with an extended secondary orbiting body that are not exact analogues of $\varepsilon$ Aurigae. Blue squares are KH 15D like systems and light blue squares are single stars with a circumstellar disk producing the eclipse. Systems that do not fall into these categories are represented with purple diamonds and green triangles are used for systems of unknown origin. ASASSN-24fw is the yellow star. The shaded areas delimit the parameter space of binary star eclipses, where eclipses deeper than 1 mag mostly lie in the light orange region. The gray dashed line is the limit where the duration is equal to the period. \label{fig:pdf_corr}}  
\end{figure*}

\begin{figure*}
\centering
\includegraphics[width=0.97\textwidth]{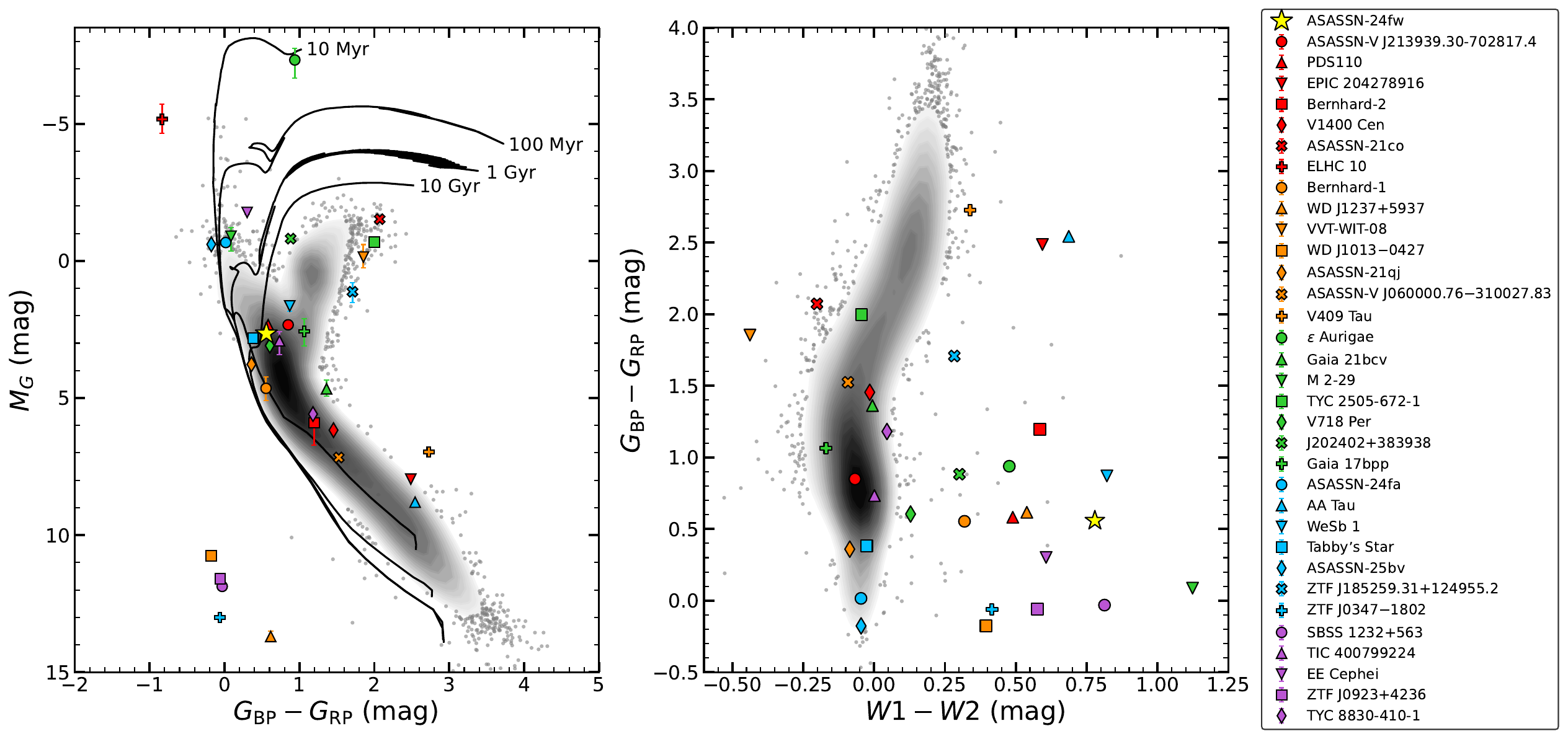}
\caption{An extinction corrected {\it Gaia} color-magnitude diagram (left) and a WISE/{\it Gaia} color-color diagram (middle) of the stars with long deep eclipses labeled as defined in the right panel. The absolute magnitude uncertainties are dominated by the distance uncertainty. The gray shaded region and points are the density distribution of 50000 randomly selected {\it Gaia} stars.  The black lines are Solar metallicity {\tt PARSEC} isochrones.\label{fig:cmd_ccd}}  
\end{figure*}

We searched the literature for similar dimming events to explore the general properties of these systems. The definition of ``long and deep dimming" is rather vague but we tried to select systems that are not normal eclipsing binaries with occultations longer than 1 day and with a flux decrease of at least 20\%. Table \ref{tab:ecl} shows the 41 selected systems ordered by the duration of the dimming. We include the $Gaia$ DR3 ID, the \citet{2021AJ....161..147B} photogeometric distance and the $E(B-V)$ extinction from {\tt mwdust} \citep{2016ApJ...818..130B} and the references for each system. The range of eclipse parameters is very broad, with durations from a few days to several years, depths ranging from a few tenths of a magnitude up to 5 magnitudes and orbital periods from days to decades. There are also aperiodic, variable or irregular eclipses in the sample. It is clear that the space of ``long and deep'' eclipses is more complex than the two archetypes, $\varepsilon$ Aurigae and KH 15D.

There are eclipses that fall into the $\varepsilon$ Aurigae category, where a star is occulted by a stellar companion surrounded by a dusty disk. This category includes $\eta$ Geminorum, Gaia17bpp, TYC 2505-672-1, ELHC 10, OGLE-LMC-ECL-17782, OGLE-LMC-ECL-11893 and OGLE-BLG182.1.162852. These eclipses tend to be smooth, without substructure during the events. However there are other systems with eclipses explained by a secondary object with an extended structure which are not twins of $\varepsilon$ Aurigae. For example, the variable duration and depth of the eclipses in EE Cephei are explained by the precession of the disk around the companion star. The companion of PDS 110 is a substellar object rather than a star, and ASASSN-21js and V1400 Cen also have a substellar companions with two and three rings around them producing irregularly shaped eclipses. Gaia21bcv is claimed to have a stellar or substellar companion with a clumpy disk that produces irregular and variable dips during the eclipse. ZTF J185259.31+124955.2 is an evolved star which is transferring mass to its companion to form the extended disk that produces the occultations. WeSb1 is a planetary nebula where a companion star collected material ejected by the former asymptotic giant branch star and this debris around the companion now produces the stochastic occultations. ASASSN-21co is an eclipsing binary formed by two M giants. And lastly, V773 Tau is a binary surrounded by a disk that occults another binary star.

There are other systems like KH 15D (binary systems with a circumbinary disk that produces the occultations), such as VSSG 26, Bernhard-1, Bernhard-2, and YLW 16A. In addition, EPIC 204278916, V409 Tau, AA Tau, V718 Per, and CHS 7797 are single stars with circumstellar disks that also produce eclipses. This type of eclipse is found preferentially in young binary or single stars. However, M 2-29 is a planetary nebula where a binary star is occulted by a disk and there is evidence for the primary being, in turn, a close binary. J202402+383938 is a Be star with a circumstellar disk which caused a 2200-day, double peaked eclipse with additional shorter events in between.

There are eclipses that cannot be assigned to these scenarios. ASASSN-21qj, Tabby’s Star, and TYC 8830-410-1 occultations are explained by debris clouds, produced by collisions between exocomets or planets around the main star. In addition, Tabby’s Star exhibits long-term variability with fading and brightening episodes from 2006 to 2017 \citep{2018ApJ...853...77S} which adds further complexity to its behavior. ZTF J0347$-$1802, SBSS 1232+563 and ZTF J0923+4236 are white dwarfs hosting debris disks that cause aperiodic and variable occultations. ZTF J0139+5245 is also a white dwarf with a debris disk but its occultations are periodic. Any periodicity for the white dwarf systems WD J1237+5937, WD J1013$–$0427 and WD J1302+1650 cannot be assessed because only one event has been detected. The dips of TIC 400799224 are proposed to be caused by dust clouds emitted by the star.

The different classes ($\varepsilon$ Aurigae like, KH 15D like, other or unknown) are presented in Figure \ref{fig:pdf_corr} in period-duration and depth-duration space. The shaded regions roughly delimit the parameter space of simple eclipsing binaries. To delimit these zones we use Solar metallicity {\tt PARSEC} isochrones and randomly select pairs of stars of the same age. Note, we are not trying to make a statistical model but simply to map the possible parameter space of stellar eclipses. We assign a period to each binary from a logarithmically uniform distribution and, assuming a circular edge-on orbit, compute the total eclipse duration and depth at $g$ band. We only keep systems where the stars are inside their binary Roche lobes using the \citet{1983ApJ...268..368E} Roche lobe approximation. Actual samples of eclipsing binaries \citep[see, e.g.,][]{2016AJ....151...68K,2022MNRAS.517.2190R,2023A&A...674A..16M}, do not fill this space, but we are interested in the extremes, not the norm. The sharp lower edge of the eclipsing binary distribution comes from the Roche radius ($t_d\sim R_\ast/v_{orb}\sim P$ for $R_\ast\sim a$, where $t_d$ is the total duration, $R_\ast$ the radius of the occulted star, $v_{orb}$ the orbital velocity, $P$ the period and $a$ the orbital radius). Systems lie along lines of $P\sim t_d^3$ as the orbital radius $a$ is varied.

A distinctive trait of the eclipses produced by companions hosting disks or other extended objects ($\varepsilon$ Aurigae like) is to have larger periods than eclipses caused by circumbinary or circumstellar disks (KH 15D like) for the same eclipse duration. Hence, the $\varepsilon$ Aurigae class spends a smaller fraction of their period in the dim phase compared to the KH 15D class. Interestingly, all KH 15D-like systems with measured periods and durations are in the region below the Roche radius of binary stars. This is only possible because their occulter completely surrounds the star (or stars), allowing for longer dimmings. However, if the geometry of ASASSN-24fw is confirmed to be like KH~15D, it will be the only known circumbinary disk lying in the allowed binary region and producing the longest eclipses with the largest period. There are no obvious patterns in the depth-duration distribution. Only a few of the long and deep eclipses are deeper than the maximum depths of binary eclipses. However, binary stars with eclipses deeper than 1 mag tend to be along the light orange strip in the period-duration figure. The long and deep eclipse population, on the other hand, is more scattered.

Most of the systems lack multiwavelength measurements to assess whether the dimming is achromatic. Of those that have multiwavelength depth measurements, Bernhard-1, Bernhard-2, Gaia21bcv, M 2-29 and WD J1013$–$0427 have chromatic dimmings but the eclipses of CHS 7797, VVV-WIT-08, $\varepsilon$ Aurigae, WD J1237+5937, WD J1302+1650, ZTF J185259.31+124955.2 and SBSS 1232+563 are achromatic. For CHS 7797, the depth is achromatic for wavelengths smaller than 2 $\mu$m, suggesting that the grain sizes of the dust material around the star are larger than usual \citep[at least 1-2 $\mu$m,][]{2012AA...544A.112R}. On the other hand, the nature of the occulter of VVV-WIT-08 is still unclear, but the $K_s$-band depth is slightly shallower than in the $I$ band. \citet{2021MNRAS.505.1992S} suggest that this can be explained if the obscuring material is completely opaque and the detected in-transit light comes from forward scattering of light by the edges of the occulter but they also note that they cannot confidently rule out large grains. The $\varepsilon$ Aurigae eclipse depth is constant in wavelength from the optical to the IR, which can be explained if the dust grains of the companion disk are larger that 10 $\mu$m \citep{2010ApJ...714..549H}, similar to our argument in Section \ref{subsec:occ}. All the white dwarfs in the sample with multiwavelength measurements have gray dimmings, except for WD J1013$–$0427. They have been modeled by a dusty debris disk made up of fairly large grains.

Figure \ref{fig:cmd_ccd} shows the systems with \citet{2021AJ....161..147B} distances in a {\it Gaia} color-magnitude diagram using the {\tt mwdust} \citep{2016ApJ...818..130B} foreground extinction estimates. We also show 50000 randomly selected {\it Gaia} stars. If specific type of stars are more prone to long and deep eclipses, the systems should be clumped in certain regions of the diagram. This is not, however, what we see in Figure \ref{fig:cmd_ccd}. Systems with long and deep eclipses populate almost all of the color-magnitude diagram. What many of them do have in common is a WISE mid-IR excess, as shown in the right panel of Figure \ref{fig:cmd_ccd}. This is not surprising given that most of the occultations are produced by dust disks and that some of the occulted stars are young stellar objects. We tried to use the longer $W3/W4$ WISE bands to look for cooler dust, but the $W3/W4$ magnitudes were dominated by systematic problems. It is likely that most of the systems that do not have mid-IR $W1/W2$ excesses have longer wavelength IR excesses given that the majority of the events are explained by dusty structures.

\section{Conclusions}\label{sec:concl}

ASASSN-24fw is a 7$L_\odot$ F type star with a mid-IR excess. Its location on the CMD is consistent with a $\sim$1.53~M$_\odot$ pre-main sequence star that could have a residual disk. A slightly evolved $\sim$1.45~M$_\odot$ star is also possible, but such stars are typically not dusty. High resolution spectra once the occultation ends may clarify the system age. If the star is young, it is likely to show Li absorption and fast rotation while, if it is old, it should not. The MODS spectrum was not sensitive enough to detect the expected Li feature  (Figure \ref{fig:spec}), but a high resolution spectrum taken with the star again bright should easily test for its presence and strength. Combining the DASCH archival photometry from 1894 to 1989 with more recent surveys and the current event, a 43.8-year period seems likely, as proposed by \citet{2024ATel16919....1N}. Unfortunately, few of the authors will live to see the next eclipse in $\sim 2068$, but some will. This also raises the question of ``slow'' astronomy \citep{2025MNRAS.539.1065P} -- since the end of the Harvard plate surveys (\citealt{2012IAUS..285...29G}), we have no project with the commitment of long term support needed to monitor the ``slow'' sky.

The star experienced a remarkably symmetric $\Delta g=4.12\pm0.02$ mag eclipse with essentially no color change, $\Delta (g-z)=0.31\pm0.15$ mag, and linearly polarized emission of up to 4\%. The deep part of the eclipse  lasted 219 days.  Large ($\sim$ 20~$\mu$m) carbonaceous or water ice grains can produce the achromaticity and polarization, while silicate and CO$_2$ grains cannot. This suggests an older, probably protoplanetary or collisional debris disk where radiation pressure has removed the smaller grains (see \citealt{2006A&A...455..509K}).  This
would leave behind the larger grains needed to produce the achromaticity and polarization. The occultation ended
just as the system set. While we do see that the polarization was decreasing as the event ended, an immediate test for the hypothesis that the occulter is producing the polarization would be to measure the polarization once the source rises again and find it to be negligible (or significantly
smaller).

We considered several possible system geometries assuming that the period of 43.8~years proposed by \cite{2024ATel16919....1N} is correct and that there is an $\sim 0.25 M_\odot$ M star secondary. The presence of such a secondary is supported by the SED and the dilution of the F star infrared absorption features near minimum, as well
as the need for something to drive the evolution of the system.  The most promising scenario is to have a 
circumbinary disk.  The 43.8~year period is
then half the precession period, assuming that events occur after every $180^\circ$ of precession.  In either
of the precession scenarios, the binary orbit would be significantly shorter than the precession time
at $P_b\sim 8(a_b/1000 R_\odot)^{3/2}$~years with an F star velocity of $\simeq 2.6(1000R_\odot/a_b)^{1/2}$~km/s
that could be constrained over a few years.

We cataloged 46 similar events to see if we could draw general conclusions about systems producing long and deep eclipses. For the $\varepsilon$ Aurigae or KH 15D archetypes we found that for the same eclipse duration, the eclipses produced by a companion hosting a disk (or extended) structure ($\varepsilon$ Aurigae like) have longer periods than those produced by disks directly surrounding the eclipsed star or stars (KH 15D like). 
The systems are scattered across an optical CMD, and there seems to be no preferred stellar type for these events. However, roughly half the systems have WISE mid-IR excesses, which is consistent with the need for obscuration to produce these events. We suspect many without $W1/W2$ excesses would show evidence of dust at longer wavelengths.\\

\section*{Acknowledgments}
We thank Annika Peter and Marshall Johnson for comments.
CSK, KZS and DMR are supported by NSF grants AST-2307385 and 2407206.
The Shappee group at the University of Hawai‘i is supported with funds from NSF (grants AST-2407205) and NASA (grants HST-GO-17087, 80NSSC24K0521, 80NSSC24K0490, 80NSSC23K1431).
The research at Boston University was supported in part by the National Science Foundation grant AST-2108622,  and NASA Fermi Guest Investigator grant 80NSSC23K1507.  EG was supported by NSF grant AST-2106927.  
This material is based upon work supported by the National Aeronautics and Space Administration under Grant No. 80NSSC23K1068 issued through the Science Mission Directorate.
We thank Las Cumbres Observatory and its staff for their continued support of ASAS-SN. ASAS-SN is funded by Gordon and Betty Moore Foundation grants GBMF5490 and GBMF10501 and the Alfred P. Sloan Foundation grant G-2021-14192.
This work made use of data from the following facilities. The MODS spectrographs built with funding from NSF grant AST-9987045 and the NSF Telescope System Instrumentation Program (TSIP), with additional funds from the Ohio Board of Regents and the Ohio State University Office of Research. The LBT is an international collaboration among institutions in the United States, Italy, and Germany. LBT Corporation partners are: The University of Arizona on behalf of the Arizona Board of Regents; Istituto Nazionale di Astrofisica, Italy; LBT Beteiligungsgesellschaft, Germany, representing the Max-Planck Society, The Leibniz Institute for Astro-physics Potsdam, and Heidelberg University; The Ohio State University, representing OSU, University of Notre Dame, University of Minnesota, and University of Virginia. The 1.83 m Perkins Telescope Observatory (PTO) in Arizona, which is owned and operated by Boston University. The Rapid Eye Mount telescope at La Silla, Chile, which is operated by the Istituto Nazionale di Astrofisica (INAF) of Italy and supported by the d'REM team.  The European Space Agency (ESA) mission {\it Gaia} (\url{https://www.cosmos.esa.int/gaia}), processed by the {\it Gaia} Data Processing and Analysis Consortium (DPAC, \url{https://www.cosmos.esa.int/web/gaia/dpac/consortium}). Funding for the DPAC has been provided by national institutions, in particular the institutions participating in the {\it Gaia} Multilateral Agreement. The Digital Access to a Sky Century @ Harvard (DASCH), which has been partially supported by NSF grants AST-0407380, AST-0909073, and AST-1313370. The Wide-field Infrared Survey Explorer, which is a joint project of the University of California, Los Angeles, and the Jet Propulsion Laboratory/California Institute of Technology, and NEOWISE, which is a project of the Jet Propulsion Laboratory/California Institute of Technology. WISE and NEOWISE are funded by the National Aeronautics and Space Administration. The TESS mission, which are publicly available from the Mikulski Archive for Space Telescopes (MAST). This work has made use of data from the Asteroid Terrestrial-impact Last Alert System (ATLAS) project. The Asteroid Terrestrial-impact Last Alert System (ATLAS) project is primarily funded to search for near earth asteroids through NASA grants NN12AR55G, 80NSSC18K0284, and 80NSSC18K1575; byproducts of the NEO search include images and catalogs from the survey area. This work was partially funded by Kepler/K2 grant J1944/80NSSC19K0112 and HST GO-15889, and STFC grants ST/T000198/1 and ST/S006109/1. The ATLAS science products have been made possible through the contributions of the University of Hawaii Institute for Astronomy, the Queen’s University Belfast, the Space Telescope Science Institute, the South African Astronomical Observatory, and The Millennium Institute of Astrophysics (MAS), Chile. The national facility capability for SkyMapper has been funded through ARC LIEF grant LE130100104 from the Australian Research Council, awarded to the University of Sydney, the Australian National University, Swinburne University of Technology, the University of Queensland, the University of Western Australia, the University of Melbourne, Curtin University of Technology, Monash University and the Australian Astronomical Observatory. SkyMapper is owned and operated by The Australian National University's Research School of Astronomy and Astrophysics. The Pan-STARRS1 Surveys (PS1) and the PS1 public science archive have been made possible through contributions by the Institute for Astronomy, the University of Hawaii, the Pan-STARRS Project Office, the Max-Planck Society and its participating institutes, the Max Planck Institute for Astronomy, Heidelberg and the Max Planck Institute for Extraterrestrial Physics, Garching, The Johns Hopkins University, Durham University, the University of Edinburgh, the Queen's University Belfast, the Harvard-Smithsonian Center for Astrophysics, the Las Cumbres Observatory Global Telescope Network Incorporated, the National Central University of Taiwan, the Space Telescope Science Institute, the National Aeronautics and Space Administration under Grant No. NNX08AR22G issued through the Planetary Science Division of the NASA Science Mission Directorate, the National Science Foundation Grant No. AST-1238877, the University of Maryland, Eotvos Lorand University (ELTE), the Los Alamos National Laboratory, and the Gordon and Betty Moore Foundation. The VizieR catalogue access tool, CDS, Strasbourg Astronomical Observatory, France (DOI : 10.26093/cds/vizier).

\bibliographystyle{aasjournal}
\bibliography{refs}{}

\end{document}